\documentclass[12pt]{article}
\pdfoutput=1
\usepackage{amsmath}
\usepackage{multirow}
\usepackage{amsfonts}
\usepackage{amssymb,graphics,psfrag}
\usepackage{array,epsfig,multirow,stmaryrd,graphicx}
\usepackage{comment}
\usepackage{slashed}
\usepackage{hyperref}
\usepackage{tensor}
\usepackage{booktabs}
\usepackage{colortbl}
\usepackage{hhline}
\usepackage{mathtools}
\usepackage{enumitem}
\usepackage{xfrac}
\usepackage{tikz}
\usepackage[nosort]{cite}
\usetikzlibrary{decorations.markings}
\usetikzlibrary{shapes,arrows}
\usetikzlibrary{decorations.pathreplacing}

\def\hybrid{\topmargin -20pt    \oddsidemargin 0pt
        \headheight 0pt \headsep 0pt 
        \textwidth 6.25in      % A4 paper
        \textheight 9 in      % A4 paper
        \marginparwidth .875in
        \parskip 5pt plus 1pt
          \jot = 1.5ex
  }
\hybrid
\numberwithin{equation}{section}
\numberwithin{table}{section}

\newcommand{\beq}{\begin{equation}}
\newcommand{\eeq}{\end{equation}}
\newcommand{\be}{\begin{equation}}
\newcommand{\ee}{\end{equation}}
\newcommand{\bea}{\begin{eqnarray}}
\newcommand{\eea}{\end{eqnarray}}
\newcommand{\ben}{\begin{eqnarray*}}
\newcommand{\een}{\end{eqnarray*}}               
\newcommand{\ba}{\begin{aligned}}
\newcommand{\ea}{\end{aligned}}
\newcommand{\bt}{\begin{tabular}}
\newcommand{\et}{\end{tabular}}
\newcommand{\bc}{\begin{center}}
\newcommand{\ec}{\end{center}}
\newcommand{\cC}{\mathcal{C}}
\newcommand{\cD}{\mathcal{D}}

\newcommand{\cK}{\mathcal{K}}
\newcommand{\cN}{\mathcal{N}}

\newcommand{\cI}{\mathcal{I}}
\newcommand{\cJ}{\mathcal{J}}
\newcommand{\cR}{\mathcal{R}}

\DeclareMathOperator{\rk}{rank}

\DeclareMathOperator{\sign}{sign}

\newcommand{\nn}{\nonumber}

\newcommand{\cref}{{\bf [check ref]}}

\newcommand{\tr}{\mathrm{tr}}

\psfrag{n1}{$\nu_1$}
\psfrag{n2}{$\nu_2$}
\psfrag{n1'}{$\nu_1'$}
\psfrag{n2'}{$\nu_2'$}
\psfrag{n9}{$\nu_9$}
\psfrag{n10'}{$\nu_{10}'$}
\psfrag{t1}{${\nu}_1$}
\psfrag{t2}{${\nu}_2$}
\psfrag{t9}{${\nu}_9$}
\psfrag{t1'}{${\nu}_1'$}
\psfrag{t2'}{${\nu}_2'$}
\psfrag{t10'}{${\nu}_{10}'$}

\newcommand{\fe}{\sfrac{1}{2}}
\newcommand{\gr}{\sfrac{3}{2}}

\newcommand{\dynkinradius}{.05cm}
\newcommand{\dynkinstep}{.25cm}
\newcommand{\dynkindot}[2]{\fill (\dynkinstep*#1,\dynkinstep*#2) circle (\dynkinradius); }

\newcommand{\dynkinline}[4]{\draw[thin] (\dynkinstep*#1,\dynkinstep*#2) -- (\dynkinstep*#3,\dynkinstep*#4);}
\newcommand{\dynkindots}[4]{\draw[dotted] (\dynkinstep*#1,\dynkinstep*#2) -- (\dynkinstep*#3,\dynkinstep*#4);}
\newcommand{\dynkindoubleline}[4]{\draw[double,postaction={decorate}] (\dynkinstep*#1,\dynkinstep*#2) -- (\dynkinstep*#3,\dynkinstep*#4);}
\newcommand{\dynkintripleline}[4]{\draw[double,postaction={decorate}] (\dynkinstep*#1,\dynkinstep*#2) -- (\dynkinstep*#3,\dynkinstep*#4);
				  \draw (\dynkinstep*#1,\dynkinstep*#2) -- (\dynkinstep*#3,\dynkinstep*#4);}

\newenvironment{dynkin}{\begin{tikzpicture}[decoration={markings,mark=at position 0.6 with {\arrow[line width=0.15mm]{>}}}]}
{\end{tikzpicture}}

\def\blfootnote{\xdef\@thefnmark{}\@footnotetext}
\long\def\symbolfootnote[#1]#2{\begingroup%
\def\thefootnote{\fnsymbol{footnote}}\footnote[#1]{#2}\endgroup}

\begin{document}

\baselineskip=15pt

\begin{titlepage}
\begin{flushright}
\parbox[t]{1.8in}{\begin{flushright}  MPP-2015-25 \end{flushright}}
\end{flushright}

\begin{center}

\vspace*{ 1.2cm}

{\Large \bf  Anomaly Cancelation in Field Theory\\[.2cm]
                   and F-theory on a Circle}

\vskip 1.2cm

\begin{center}
 {Thomas W.~Grimm and Andreas Kapfer\ \footnote{grimm,\ kapfer@mpp.mpg.de}}
\end{center}
\vskip .2cm
\renewcommand{\thefootnote}{\arabic{footnote}}

{Max-Planck-Institut f\"ur Physik, \\
F\"ohringer Ring 6, 80805 Munich, Germany}

 \vspace*{1cm}

\end{center}

\vskip 0.2cm
 
 \begin{center} {\bf ABSTRACT } \end{center}

We study the manifestation of local gauge anomalies of four- and
six-dimensional field theories in the lower-dimensional Kaluza-Klein theory
obtained after circle compactification. We identify a convenient set of transformations 
acting on the whole tower of massless and massive states and investigate 
their action on the low-energy effective theories in the Coulomb branch.
The maps employ higher-dimensional large
gauge transformations and precisely yield the anomaly cancelation conditions when 
acting on the one-loop induced Chern-Simons terms in the 
three- and five-dimensional effective theory. The arising symmetries are argued to play a key role in 
the study of the M-theory to F-theory limit on Calabi-Yau manifolds. For example, using the fact that
all fully resolved F-theory geometries inducing multiple Abelian gauge groups or non-Abelian 
groups admit a certain set of symmetries, we are able to generally show the cancelation
of pure Abelian or pure non-Abelian anomalies in these models.

\vskip 0.4cm

\hfill {February, 2015}
\end{titlepage}

\tableofcontents

\newpage

%%%%%%%%%%%%%%%%%%

%%%%%%%%%%%%%%%%%%%%%%%%%%%%%%%%%%%%%%%%%%

\section{Introduction}

A study of the consistency of quantum field theories requires to investigate 
their local symmetries both at the classical and quantum level. 
In particular, even if such gauge symmetries are manifest in the classical 
theory, they might be broken at the quantum level and induce
a violation of essential current conservation laws. Such 
inconsistencies manifest themselves already
at one-loop level and are known as anomalies \cite{ZinnJustin:2002vj,Harvey:2005it,Bilal:2008qx}. 
Four-dimensional quantum field theories with chiral spin-$\sfrac{1}{2}$ fermions, for example, 
can admit anomalies which signal the breaking of the gauge symmetry. 
Consistency requires the cancelation of these anomalies either by 
restricting the chiral spectrum such that a cancelation among various 
contributions takes place, or by implementing a generalized 
Green-Schwarz mechanism \cite{Green:1984sg,Sagnotti:1992qw}. 
The latter mechanism requires the presence of a 
$U(1)$ gauged axion-like scalar with tree-level diagrams 
canceling the one-loop anomalies. In six space-time dimensions 
anomalies pose even stronger constraints, since in addition to 
spin-$\sfrac12$ fermions also spin-$\sfrac{3}{2}$ and two-tensors can be 
chiral. Also in this case a generalized Green-Schwarz mechanism can 
be applied to cancel some of these anomalies.

In this work we address the manifestation of anomaly cancelation 
in four-dimensional and six-dimensional field theories from a Kaluza-Klein perspective 
when considering the theories to be compactified on a circle. 
Note that on a circle one can expand all higher-dimensional fields
into Kaluza-Klein modes yielding a massless lowest mode and 
a tower of massive excited modes. 
Clearly, keeping track of this infinite set of fields one retains the full information 
about the higher-dimensional theory, including its anomalies. 
In a next step, one can compute the lower-dimensional effective 
theory for the massless modes only. This requires to integrate out 
all massive states. Of particular interest for the discussion of 
anomalies are the effective lower-dimensional couplings   
that are topological in nature. These do not continuously depend on 
the cutoff scale and therefore might receive quantum corrections 
from integrating out the massive states. Prominent examples are 
three-dimensional gauge Chern-Simons 
terms as well as five-dimensional 
gauge and gravitational Chern-Simons terms.
These couplings are indeed modified at one-loop when integrating out 
massive states. In three dimensions only certain massive spin-$\sfrac12$ 
fermions contribute \cite{Niemi:1983rq,Redlich:1983dv,Aharony:1997bx}, while in five dimensions also massive spin-$\sfrac{3}{2}$ and 
massive self-dual tensors give a non-vanishing shift \cite{Witten:1996qb,Bonetti:2012fn,Bonetti:2013ela}. In fact, precisely 
those modes contribute that arise from higher-dimensional chiral fields. Therefore, 
one expects that the Chern-Simons terms of the effective theories encode 
information about the higher-dimensional anomalies. This was recently investigated 
motivated by the study of F-theory effective actions via M-theory in \cite{Grimm:2011fx,Cvetic:2012xn,Bonetti:2011mw,Grimm:2013oga,Cvetic:2013uta,Anderson:2014yva}. With a different motivation similar questions were 
addressed in \cite{Landsteiner:2011iq,Loganayagam:2012pz,Golkar:2012kb,Landsteiner:2012kd,Jensen:2012kj,Jensen:2013kka,Jensen:2013rga,DiPietro:2014bca} in the study of applications of holography. 

The connection between one-loop Chern-Simons terms and 
anomalies in the higher-dimensional theory, while expected to exist, was only shown to be rather indirect. In fact, 
it is not at all obvious how the anomaly cancelation conditions arise, for example, 
from comparing classical and one-loop Chern-Simons terms. While for many concrete 
examples in F-theory it was possible to check anomaly cancelation using the lower-dimensional 
effective theory and Chern-Simons terms arising from M-theory, there was no known systematics behind this 
as of now. In this work we will suggest that there is an elegant way to actually approach 
this generally by describing symmetry transformations among effective theories that 
exist if higher-dimensional anomalies are canceled.

Let us consider an effective theory obtained after circle reduction. If the higher-dimensional 
theory admits a gauge group one can use the Wilson-line scalars of the gauge fields 
around the circle to move to the lower-dimensional Coulomb branch. In other words, 
one considers situations in which these Wilson line scalars admit a vacuum expectation 
value, which we call Coulomb branch parameters in the following. 
The masses of all the massive states are now dependent both on the circle radius, if they 
are excited Kaluza-Klein states,  
and on the Coulomb branch parameters, if they where charged under the higher-dimensional gauge group. 
With this in mind one can then compute the effective theory 
for the massless modes and focus on the Chern-Simons terms. While the one-loop
Chern-Simons terms are not continuous functions of the masses of the integrated-out 
states, they can experience discrete shifts, when changing the radius or the 
Coulomb branch parameters. In other words, depending on the background value 
of the Coulomb branch parameters and the radius, the effective theories can take 
a different form. A priori, one would thus expect that one finds infinitely many values for the 
Chern-Simons coefficient due to the infinitely many hierarchies of Kaluza-Klein masses
and Coulomb branch masses. However, we argue that there are certain
transformations arising from higher-dimensional large gauge transformations 
that identify different Coulomb branch parameters and effective theories if 
and only if anomalies are canceled. Importantly, the transformations 
are designed to yield the anomaly cancelation conditions when considering the 
classical and one-loop Chern-Simons terms of the effective theory.
Our goal is to examine these maps both for four-dimensional and 
six-dimensional gauge theories with a focus on 
pure non-Abelian and pure Abelian anomalies. 

While most of our discussion is purely field-theoretic, it is important to 
stress that the original motivation to carry out such a study arose 
from the analysis of anomaly cancelation in F-theory via the M-theory dual.
Recall that F-theory on a complex four- or three-dimensional manifold 
yields a four-dimensional and six-dimensional effective theory, respectively \cite{Vafa:1996xn,Morrison:1996na,Morrison:1996pp,Ferrara:1996wv,Denef:2008wq,Grimm:2010ks,Bonetti:2011mw,Grimm:2013oga}.
The geometry of the internal manifold dictates both the gauge group and 
matter content that arises from space-time filling seven-branes. To study four- and 
six-dimensional field theories that arise from F-theory, however, one needs to take 
a detour via M-theory. In fact, one can derive an effective theory in three or 
five dimensions by starting with eleven-dimensional supergravity dimensionally reduced on 
the completely resolved geometry. Different gauge theory phases of such theories and 
their relation to geometric resolutions have been recently studied in \cite{Intriligator:1997pq,Grimm:2011fx,Hayashi:2013lra,Hayashi:2014kca,Esole:2014bka,Esole:2014hya,Braun:2014kla}.
These effective theories are in the lower-dimensional 
Coulomb branch and dual to the F-theory effective actions on an extra circle with all massive 
modes integrated out. Therefore, one is precisely in the situation we consider in 
purely field-theoretic terms. 

The transformations we consider in order 
to map effective theories descend from transformations acting on the 
resolved F-theory geometries. This implies that if the transformations 
are in fact symmetries of the geometries and the M-theory to F-theory limit,
then anomalies are canceled.  Indeed we are able to show that in purely Abelian theories the 
transformations on the geometry correspond to \textit{picking a zero-section},
i.e.~identifying the Kaluza-Klein vector of the F-theory circle compactification 
in the internal geometry. Since nothing dictates a preferred choice of 
zero-section this has to be a symmetry of the M-theory to F-theory 
limit. This shows that anomalies for purely Abelian theories are generally 
canceled for the considered F-theory geometries. In order to give 
a proof of generic cancelation of non-Abelian anomalies we need 
to identify the corresponding geometrical symmetry. Remarkably, 
we find that it corresponds to `\textit{picking a zero-node}', which was as of 
now always done by a canonical choice using the zero-section. 
Having established the presence of this extra symmetry we argue for the
generic cancelation of pure non-Abelian anomalies in the 
considered F-theory geometries. 

This paper is organized as follows. In \autoref{sec:4d_anomalies} and \autoref{sec:6d_anomalies}
we perform the field-theoretic analysis of anomalies in four and six 
dimensions, respectively. In both dimensions we first focus on purely non-Abelian 
models and later on purely Abelian models. We give a detailed account 
of the transformations that become symmetries among effective theories 
once anomalies are canceled. In \autoref{sec:F-theory} we turn to the 
analysis of the F-theory geometries and make extensive use 
of the duality between M-theory and F-theory on an extra circle. 
Identifying actual symmetries of the geometries we are able 
to show the general cancelation of pure Abelian and pure non-Abelian 
anomalies. We supplement additional information in four appendices. 
Relevant identities valid on the Coulomb branch are discussed in 
\autoref{CB_id}, while details on the one-loop computations are 
summarized in \autoref{app:loops}.
Useful group theory identities are summarized in \autoref{app:traces},
where we also translate trace identities into relations among weights.
Some important results on the intersection numbers of our geometries 
are given in \autoref{sec:intersection_nos}.

\section{Anomalies of four-dimensional theories}\label{sec:4d_anomalies}

In this section we study anomaly cancelation of four-dimensional matter-coupled 
gauge theories from the three-dimensional Kaluza-Klein perspective of the circle 
compactified theories on the Coulomb branch. More precisely, we analyze the 
Chern-Simons terms of classes of three-dimensional effective theories that arise at different 
values of the Coulomb branch parameters after integrating out all massive modes. 
After introducing the generalities on the four-dimensional setup in \autoref{general_setup4D},
we discuss the circle reduction and one-loop Chern-Simons terms in \autoref{sec:3DCSterms}.
If anomalies are canceled, a set of transformations induced by a higher-dimensional 
large gauge transformation identifies infinitely many effective theories
and thus represents a symmetry.
%Remarkably these symmetries exist if and only if anomalies are canceled in the four-dimensional theory,
%since the transformation of one-loop Chern-Simons terms in the three-dimensional theory precisely reproduces the anomaly cancelation conditions.
The motivation to study these transformation arose originally from the duality of M-theory to F-theory, 
as detailed in \autoref{sec:F-theory}, but applies to arbitrary matter-coupled gauge theories in four dimensions, 
including also a possible coupling to gravity. For simplicity we will
only investigate the two cases of a pure non-Abelian and pure Abelian gauge group in \autoref{sec:4d_nonAb} and \autoref{sec:4d_Ab}, respectively. We are confident that the reasoning generalizes also to reductive gauge groups.

\subsection{General setup and anomaly conditions} \label{general_setup4D}

Let us first introduce the general setup before restricting to pure non-Abelian and pure Abelian gauge groups, respectively.
We use $G$ to denote a simple gauge group\footnote{The generalization to semi-simple gauge groups is straightforward.}
with gauge bosons $\hat A$. Denoting by $T_\cI$, $\cI = 1,\ldots,\text{dim}\,G$ the Lie algebra generators, we expand 
\beq
   \hat A = \hat A^\cI T_\cI  = \hat A^I T_I + \hat A^{\boldsymbol{\alpha}} T_{\boldsymbol{\alpha} }
\eeq
where $T_I$, $I = 1 ,\dots ,\rk G$ are the generators of the Cartan subalgebra and $T_{\boldsymbol \alpha}$
are the remaining generators labeled by the roots ${\boldsymbol \alpha}$. 
Our conventions in the theory of Lie algebras and their 
representations are listed in \autoref{app:traces}.
The $n_{U(1)}$ Abelian gauge bosons are denoted by $\hat A^m$ with $m =1,\dots ,n_{U(1)}$.

Our setup also includes charged matter and we focus on chiral  spin-$\sfrac{1}{2}$ Weyl fermions 
since they are the only states contributing gauge anomalies. We write $F(R)$ for the chiral index of a representation $R$ of these fermions $\hat\psi$ under 
the gauge group $G$. Their
covariant derivative reads 
\begin{align} \label{def-DpsinA}
 \cD_\mu \hat\psi = \nabla_\mu \hat\psi - i \hat A^\cI_\mu  T^R_\cI \ \hat\psi \, ,
\end{align}
where $T^R_\cI $ are the Lie algebra generators in the representation $R$. Upon choosing
an eigenbasis associated to the weights $w$ of $R$ and expanding $\hat \psi$ 
with coefficients $\hat \psi(w)$  one finds for the Cartan directions 
\beq \label{wI-def}
  T^R_I \ \hat\psi(w) = w_I \ \hat \psi(w)\ , \qquad 
   w_I := \langle \boldsymbol{\alpha}^\vee_I , w \rangle \, ,
\eeq
where $\boldsymbol{\alpha}^\vee_I$ is the simple coroot associated to $T_I$.
Similarly, $F(q)$ is the chiral index of spin-$\sfrac{1}{2}$ Weyl fermions $\hat\psi$ with charges $q =(q_m)$ under $\hat A^m$
encoded in the covariant derivative
\begin{align} \label{def-DpsiA}
 \cD_\mu \hat\psi = \nabla_\mu \hat\psi - i q_m \hat A^m_\mu \ \hat\psi \, .
\end{align}
Moreover, we restrict to fermions without four-dimensional mass terms.
Note that since we treat non-Abelian and Abelian theories separately, we do not need fermions that
are charged both under $\hat A$ and $\hat A^m$.

In order to cancel anomalies a four-dimensional Green-Schwarz mechanism
might be required \cite{Green:1984sg,ZinnJustin:2002vj,Harvey:2005it,Bilal:2008qx}. Therefore, we also allow for a number of $n_{\rm ax}$ axions 
$\hat\rho_\alpha$, $\alpha =1,\dots ,n_{\rm ax}$ with covariant derivative
\begin{align}\label{e:axion_gauging}
 \cD \hat\rho_\alpha = d  \hat\rho_\alpha + \theta_{\alpha m} \hat A^m
\end{align}
with $\theta_{\alpha m}$ constant. The possibly gauge non-invariant couplings of the 
axions to the gauge fields read 
\begin{align}\label{e:4d_GS}
 \hat S_{\textrm{GS}} = \int -\frac{1}{4} b^\alpha_{mn}  \hat\rho_\alpha \ \hat F^m \wedge \hat F^n
  -\frac{1}{4\ \lambda(G)} b^\alpha \hat\rho_\alpha \ \tr_f \hat F \wedge \hat F   \, ,
\end{align}
where $b^\alpha_{mn}$, $b^\alpha$ are the Green-Schwarz coefficients, the trace $ \tr_f $
is in the fundamental representation of $G$, and $\lambda(G)$ are normalization 
factors discussed in \autoref{app:traces}.\footnote{In gravity theories there may be an additional Green-Schwarz coupling in order to cancel mixed gravitational anomalies. Since we only treat gauge anomalies in four dimensions, 
we will omit this term in the following. Also the Green-Schwarz coupling for the non-Abelian gauge fields $b^\alpha$ only plays a role in the cancelation of
mixed Abelian-non-Abelian anomalies that are not discussed in this paper. We nevertheless list it here because it will appear
when treating the M-/F-theory duality in \autoref{sec:F-theory}.}
We stress that our considerations work both in theories 
with or without an implemented Green-Schwarz mechanism. 
Furthermore, it is of course possible to include other non-chiral or 
uncharged fields, since these contribute neither to the anomalies nor to the one-loop
Chern-Simons terms of the circle-reduced theory discussed in \autoref{sec:3DCSterms}.
Finally, it is also possible to couple the theory to gravity. The arguments 
concerning the gauge anomalies will be unaltered by this generalization.

In \autoref{sec:4d_nonAb} and \autoref{sec:4d_Ab} 
we will see that via a certain basis transformation of the vectors one can recover the four-dimensional
gauge anomaly equations from one-loop Chern-Simons terms in three dimensions.
Let us therefore display the purely non-Abelian and purely Abelian anomaly conditions in four dimensions \footnote{All
symmetrizations over $n$ indices include a factor of $\frac{1}{n!}$.} 
\begin{subequations}
\begin{align}
 \label{4d_nA_anomaly}\sum_R F(R) \, V_R &= 0\ , \\
 \label{4d_A_anomaly}\sum_q F(q) \, q_m q_n q_p &= \frac{3}{2} b^\alpha_{(mn}\theta_{p)\alpha} \, ,
\end{align}
\end{subequations}
where $V_R$ is defined as
\begin{align}
 \tr_R \hat F^3 = V_R\, \tr_f \hat F^3 \, .
\end{align}
Here $\tr_R$, $\tr_f$ are the traces in the representation $R$ and the fundamental representation respectively.
We have now introduced all relevant parts of the field theory setup that are necessary for our considerations.

\subsection{Circle compactification and Chern-Simons terms} \label{sec:3DCSterms}

In the next step we compactify this theory on a circle and push it to the Coulomb branch
by allowing for a non-vanishing Wilson line background of the gauge fields reduced on the 
circle. 

We indicate four-dimensional quantities by a hat, while three-dimensional ones lack a hat.
At lowest Kaluza-Klein (KK)  level we get $\text{dim}\, G$ $U(1)$ gauge fields $A^\cI$ and
$\text{dim}\, G$ Wilson line scalars $\zeta^\cI$ from reducing $\hat A$. 
%, 
%the W-bosons become massive on the Coulomb branch.
In the Abelian sector we obtain $n_{U(1)}$ $U(1)$ gauge fields $A^m$ and Wilson line scalars $\zeta^m$ 
from reducing $\hat A^m$.
Furthermore, there are $n_{\rm ax}$ Abelian gauge fields $A^\alpha$ from dualizing the reduced $\hat\rho^\alpha$.
From the four-dimensional metric one finds at lowest level the three-dimensional metric 
$g_{\mu \nu}$, the KK-vector $A^0$ and the radius $r$ of the circle.
More precisely, we expand the four-dimensional metric as
\begin{align}\label{e:metric_reduction}
 d \hat s^2 = g_{\mu\nu}dx^\mu dx^\nu + r^2 Dy^2 \, , \qquad Dy := dy - A^0_\mu dx^\mu \, ,
\end{align}
with $x^\mu$ the three-dimensional coordinates and $y$ the coordinate along the circle.
The three-dimensional metric $g_{\mu\nu}$ is taken to be that of Minkowski space for simplicity. 
The gauge fields are expanded according to
\begin{align} \label{red-AI}
 \hat A^\cI = A^\cI - \zeta^\cI r Dy \, , \qquad \hat A^m = A^m - \zeta^m r Dy \, .
\end{align}
Note that the $\zeta^\cI$ are transforming in the adjoint representation of $G$. 

The massive fields in the theory are the KK-modes and states that acquire masses on the Coulomb branch.
The three-dimensional Coulomb branch is parametrized by the background values of the scalars $\zeta^\cI$
and $\zeta^m$, by setting 
\beq
  \langle \zeta^I \rangle \neq 0 \ , \qquad \langle \zeta^{\boldsymbol \alpha} \rangle = 0 \ , \qquad  \langle \zeta^m \rangle \neq 0 \ ,
\eeq
i.e.~giving the Cartan Wilson line scalars a vacuum expectation value (VEV). This breaks $G \times U(1)^{n_{U(1)}} \rightarrow U(1)^{\rk G} \times U(1)^{n_{U(1)}} $ and 
gives the W-bosons $A^{\boldsymbol \alpha}$ a mass. Also the modes of the higher-dimensional 
charged matter states will gain a mass. 
To find the Coulomb branch masses $m_{\rm CB}^w$ for a state $\psi(w)$ labeled 
with a weight $w$, or a state $\psi(q)$ with charge $q$, we use \eqref{def-DpsinA}-\eqref{def-DpsiA} to read off 
\beq
   m_{\rm CB}^w = w_I  \langle \zeta^I \rangle\ , \qquad m_{\rm CB}^q = q_m  \langle \zeta^m \rangle\ ,
\eeq 
In total the mass of a field $\psi_{(n)}(w)$ at KK-level $n$ in the
three-dimensional theory reads
\begin{align}\label{e:KK-masses}
 m = \begin{cases}
      m^w_{\rm CB} + n \,m_{\rm KK} = w_I   \langle\zeta^I \rangle + \frac{n}{\langle r\rangle}, \qquad &\textrm{non-Abelian gauge group}\, , \\[.2cm]
      m^q_{\rm CB} + n \,m_{\rm KK} = q_m  \langle\zeta^m \rangle + \frac{n}{\langle r\rangle},  &\textrm{Abelian gauge group}  \, ,
     \end{cases}
\end{align}
with $m_{\rm KK} = 1/\langle r \rangle$ being the unit KK-mass determined by the background value of the radius. 

Of key importance in this paper are the three-dimensional Chern-Simons terms. For a general Abelian 
theory they take the form 
\begin{align}\label{e:def_CS_4d}
 S_{\textrm{CS}} = \int \Theta_{\Lambda \Sigma} A^\Lambda \wedge F^\Sigma \, ,
\end{align}
where $\Theta_{\Lambda \Sigma} $ are constants.
We introduced the collective index $\Lambda = (0,I,\alpha)$ or $\Lambda = (0,m,\alpha)$, respectively.
Performing a classical Kaluza-Klein reduction we straightforwardly find the Chern-Simons coefficients
\begin{align}\label{e:CS_class_4d}
 \Theta_{\alpha\beta} = 0 \, , \qquad 
 \Theta_{\alpha 0} = 0 \, , \qquad 
 \Theta_{\alpha I} = 0 \, , \qquad 
 \Theta_{\alpha m} = \frac{1}{2}\theta_{\alpha m} \, .
\end{align}
Note that the non-zero coefficient descends from the Green-Schwarz couplings in four dimensions.

We are interested in the three-dimensional effective theory for the massless modes only. This implies 
that all massive states need to be integrated out. Importantly, one thus needs to 
include one-loop corrections to the Chern-Simons terms.
It is well-known that a massive charged spin-$\sfrac12$-fermion
contributes to $\Theta_{\Lambda\Sigma}$ as \cite{Niemi:1983rq,Redlich:1983dv,Aharony:1997bx}
\begin{align}\label{e:4d_single_loop_CS}
 \Theta_{\Lambda\Sigma}^{\textrm{loop}} = \frac{1}{2} q_\Lambda q_\Sigma \sign (m) \, ,
\end{align}
where $q_\Lambda$ is the charge of the fermion under the $U(1)$ gauge boson $A^\Lambda$ and the sign of the mass encodes the
spinor representation of the fermion.
We adopt the convention that a KK-mode $\psi_{(n)}$ is charged under the KK-vector in the following way
\begin{align}\label{e:KK_charge_convention}
 D_\mu \psi_{(n)} = \partial_\mu \psi_{(n)} + i n A_\mu^0 \psi_{(n)} 
\end{align}
This means that the charge of $\psi_{(n)}$ under $A^0$ is $q_0 = -n$. 

Summing up all contributions to the one-loop Chern-Simons terms 
in the circle-reduced theory requires to include the infinite KK-tower
that needs to be treated with zeta function regularization.
The relevant computations are carried out in \autoref{app:loops}.
For the present setup the relevant total one-loop Chern-Simons
coefficients for the pure non-Abelian and the pure Abelian theory respectively are evaluated as \cite{Grimm:2013oga,Cvetic:2013uta}
\begin{subequations}
\begin{align}
 \label{e:4d_nA_regular3}\Theta_{IJ} &=  \sum_{R} F(R) \sum_{w \in R}  \ (l_w +\frac{1}{2} ) \ w_I w_J \ \sign  (m^w_{\rm CB}) \, , \\
  \label{e:4d_A_regular3}\Theta_{mn} &=  \sum_{q}  F(q) \ (l_q +\frac{1}{2} ) \ q_m q_n \ \sign  (m^q_{\rm CB}) \, ,
\end{align}
\end{subequations}
where the sums in the first equation are over all representations $R$ of $G$ and all weights of a given representation.
We stress that in all sums over weights in this paper the multiplicity factors of the weights are always meant to be implicitly
included into the sum.  
The integers $l_w$, $l_q$ are defined as
\begin{align}\label{e:def_lq}
 l_w = \bigg\lfloor \Big\vert \frac{m^w_{\rm CB}}{m_{\rm KK}} \Big\vert \bigg\rfloor \, , \qquad
 l_q = \bigg\lfloor \Big\vert \frac{m^q_{\rm CB}}{m_{\rm KK}} \Big\vert \bigg\rfloor \, ,
\end{align}
where the brackets indicate the use of the floor function. The $l_w$, $l_q$ indicate at 
which KK-level the sign of the total mass of a state becomes independent of the Coulomb branch mass. 
The remaining one-loop Chern-Simons coefficients are listed for completeness in \autoref{app:loops}.
We stress that the classical Chern-Simons coefficients \eqref{e:CS_class_4d} receive no corrections at one-loop.

Having introduced the relevant parts of the circle-reduced effective theory,
we are now in a position to show in detail how gauge anomalies in four dimensions are related to symmetries on the Coulomb branch in three dimensions.

\subsection{Non-Abelian models} \label{sec:4d_nonAb}

In this section we drop the Abelian gauge bosons $\hat A^m$ and restrict to a pure non-Abelian simple gauge group $G$.
We consider a certain basis transformation of the vectors in the circle reduced theory.\footnote{
This transformation is inspired by the M-/F-theory duality, as explained in \autoref{sec:F-theory}, but our 
analysis is purely field-theoretic in the following.}
Firstly, choose an arbitrary gauge field $A^{\tilde 0}$ out of
the Cartan vectors $A^I$. The remaining Cartan vectors are denoted by $A^{\tilde I}$, while the remaining gauge fields 
of the whole set $A^\cI$, $\cI = 1,\ldots, \text{dim}\,G$ will be denoted by $A^{\tilde \cI}$.
Then we define the transformation
\begin{align}
\label{e:non_abelian_trafo_vectors_4d}
 \begin{pmatrix}
  \tilde A^{0}\\[8pt]
  \tilde A^{\tilde 0}\\[8pt]
  \tilde A^{\tilde \cI}\\[8pt]
  \tilde A^{\alpha}
 \end{pmatrix} = 
\begin{pmatrix*}[r]
 1 & 0 & 0 & 0 \\[8pt]
 1 & 1 & 0 & 0 \\[8pt]
 0 & 0 & \delta^{\tilde \cI}_{\tilde \cJ} & 0 \\[8pt]
 0 & 0 &
 0 & \delta_\beta^\alpha
\end{pmatrix*}\cdot
\begin{pmatrix}
  A^{0}\\[8pt]
  A^{\tilde 0}\\[8pt]
  A^{\tilde \cJ}\\[8pt]
  A^{\beta} \, 
 \end{pmatrix}.
\end{align}
All quantities in the transformed basis are labeled by a tilde.
In the following we again collectively denote the fields $\tilde A^{\tilde 0},\tilde A^{\tilde \cI}$ by $\tilde A^{\cI}$.
It is essential to notice that the KK-vector mixes with the Cartan $U(1)$ gauge field $A^{\tilde 0}$,
since this fact will render the basis transformation non-trivial.
This implies that, as long as one keeps all gauge fields including the W-bosons, the non-Abelian gauge 
transformations are realized on the tilted basis in a very non-trivial fashion. 
It is also worthwhile to note that the map leaves the classical Chern-Simons terms \eqref{e:CS_class_4d} invariant.

The transformation \eqref{e:non_abelian_trafo_vectors_4d} also 
requires to transform the charged fields in the KK-theory. 
Given a matter state $\psi_{(n)}$ at KK-level $n$ in the representation $R$ of $G$
one first chooses a basis associated to the weights writing $\psi_{(n)}(w)$ 
as in \eqref{wI-def}.
The transformation \eqref{e:non_abelian_trafo_vectors_4d} mixes these states as
 \begin{align}
\label{e:non_abelian_trafo_charges_4d}
\psi_{(n)}(w) \rightarrow \psi_{(\tilde n)}(\tilde w) \ ,\qquad \begin{pmatrix}
  \tilde n\\[8pt]
  \tilde w_{\tilde 0}\\[8pt]
  \tilde w_{\tilde I}
 \end{pmatrix} = 
\begin{pmatrix*}[r]
 1 & 1 & 0  \\[8pt]
 0 & 1 & 0  \\[8pt]
 0 & 0 & 1 
\end{pmatrix*}\cdot
\begin{pmatrix}
  n\\[8pt]
  w_{\tilde 0}\\[8pt]
  w_{\tilde I}
 \end{pmatrix} \, .
\end{align}
Note that in general this transformation shifts the whole KK-tower, but there is still no state charged under $\tilde A^{\alpha}$.
As we will see momentarily, the shift \eqref{e:non_abelian_trafo_charges_4d} has important consequences due to the fact that the contributions of the infinite KK-tower to the Chern-Simons terms have to be regularized.  

Let us now investigate the impact of the transformations \eqref{e:non_abelian_trafo_vectors_4d} and \eqref{e:non_abelian_trafo_charges_4d} on the effective theories for the massless fields only. 
We stress that there is not only one effective theory, but rather an infinite set of such theories 
labeled by the vacuum expectation values $\langle \zeta^I \rangle$, $\langle r \rangle$
that control the masses of the states. The effective theories are distinguished, in particular, through 
their one-loop Chern-Simons terms \eqref{e:4d_nA_regular3} and \eqref{e:4d_A_regular3}, 
which change when changing  $\langle \zeta^I \rangle$, $\langle r \rangle$.  
Note that compatibility with \eqref{red-AI} requires that 
the transformation \eqref{e:non_abelian_trafo_vectors_4d}  is accompanied by a shift 
of these vacuum expectation values as 
\begin{align} \label{shift-vevs}
 \frac{1}{\langle r \rangle} \mapsto \frac{1}{\langle r \rangle} \, , \qquad
 \langle\zeta^{\tilde I}\rangle \mapsto \langle\zeta^{\tilde I} \rangle\, , \qquad
 \langle\zeta^{\tilde 0} \rangle\mapsto \langle\zeta^{\tilde 0}\rangle - \frac{1}{\langle r \rangle} \, .
\end{align}
In other words, such a transformation in general maps one effective theory 
to a  different effective theory. However, note that it is not hard to check that \eqref{shift-vevs}
can be understood as a large gauge transformation in the underlying four-dimensional theory. 
Of course, if the theory is gauge-invariant the two effective theories related by this transformation 
need to be identical. Gauge-invariance is here tested at the loop level since we consider one-loop
Chern-Simons terms.  
Indeed, we will show in the following that the transformations \eqref{e:non_abelian_trafo_vectors_4d}, \eqref{e:non_abelian_trafo_charges_4d}, and \eqref{shift-vevs} 
identify two three-dimensional theories with distinct Coulomb branch parameters but
equivalent effective theories, if and only if four-dimensional anomalies are canceled.

To make this more precise we first perform the map \eqref{e:non_abelian_trafo_vectors_4d}
to the tilted basis. In this transformed basis we directly evaluate the one-loop
Chern-Simons coefficient $\tilde \Theta_{IJ}$ using \eqref{e:4d_nA_regular3} as
\begin{align}\label{e:4D_na_direct1}
 \tilde \Theta_{I J} = \sum_{R} F(R)\sum_{w \in R}\,w_{I} w_{J}\, \big(\tilde l_{w}+\frac{1}{2}\big)\, 
 \sign (\tilde m_{\rm CB}^{w})\, ;
\end{align}
note that $\tilde w_{I} = w_{I}$.
Alternatively we can transform $\tilde \Theta_{IJ}$ back to the old basis and perform the loop-calculation there
\begin{align}\label{e:4D_na_indirect1}
 \tilde \Theta_{I J} &= \Theta_{I J} 
= \sum_{R} F(R) \sum_{w \in R} \, w_{I} w_{J}\,\big(l_{w}+\frac{1}{2})\, \sign (m_{\rm CB}^{w}\big) \, .
\end{align}
We can now match \eqref{e:4D_na_direct1} with \eqref{e:4D_na_indirect1} and use the identity
\begin{align}\label{e:weight_id}
 w_{\tilde 0} &= \big(l_{w} + \frac{1}{2}\big) \sign (m^{w}_{\rm CB}) - \big(\tilde l_w + \frac{1}{2}\big) \sign (\tilde m^w_{\rm CB}) \, ,
\end{align}
which is proven in \autoref{CB_id}. This yields the equation
\begin{align}\label{e:non_Abelian_anomaly_4d}
\sum_{R} F(R)\sum_{w \in R}\,w_{I} w_{J} w_{\tilde 0} = 0 \, ,
\end{align}
which not only implies the pure non-Abelian anomaly \eqref{4d_nA_anomaly} but is actually equivalent to it, since for some
combinations of indices already the sum $\sum_{w \in R}\,w_{I} w_{J} w_{\tilde 0}$ is trivially zero as a group theoretical
consequence. This is shown in \autoref{app:traces}.
Performing this matching for the other one-loop Chern-Simons coefficients $\tilde\Theta_{00}$, $\tilde\Theta_{0I}$
and picking different gauge fields for the distinguished field $A^{\tilde 0}$ one again obtains either
the anomaly condition or trivially zero as consequence of representation theory.

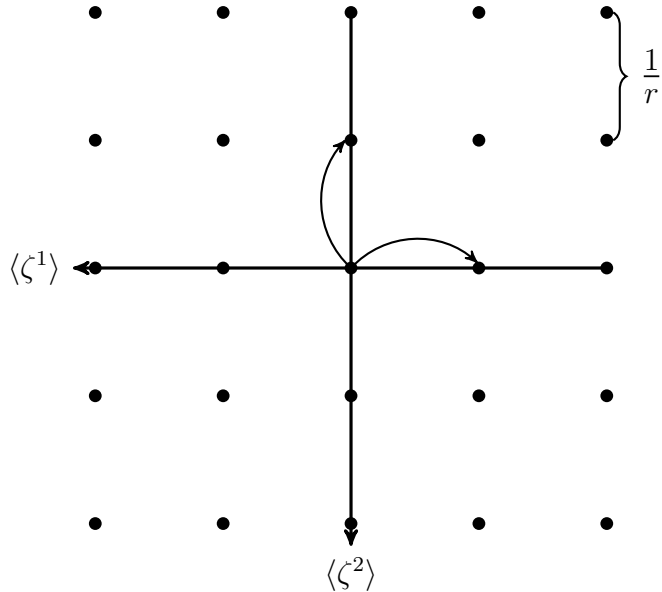
\begin{figure}
\begin{center}
\begin{tikzpicture}[scale = 0.85, axis/.style={very thick, ->, >=stealth'}]
 \foreach \x in {1,...,5} 
  { \foreach \y in {1,...,5} 
      {\fill (\x*2cm,\y*2cm) circle (0.1cm); }}
 \draw[axis]  (10cm,6cm) -- (1.65cm,6cm) node(xline)[left] {$\langle\zeta^1 \rangle$};
 \draw[axis]  (6cm,10cm) -- (6cm,1.65cm) node(yline)[below] {$\langle\zeta^2 \rangle$};
 \path[thick, ->, >=stealth'] (6cm,6cm) edge [bend left=45] (7.98cm,6.08cm);
 \path[thick, ->, >=stealth'] (6cm,6cm) edge [bend left=45] (5.91cm,8.00cm);
 \draw[thick,decorate,decoration={brace,amplitude=5pt}] (10.1cm,10cm) -- (10.1cm,8cm);
 \node at (10.7cm,9cm) {\Large $\frac{1}{r}$};
\end{tikzpicture}
\end{center}
\caption{The transformations \eqref{e:non_abelian_trafo_vectors_4d} generate a $\mathbb{Z}^{\rk G}$ group
relating theories with different Coulomb branch parameters
but the same effective theory, as long as anomalies are canceled in the four-dimensional theory. We display the lattice of related theories
for $\rk G =2$. The transformation associated to the horizontal arrow corresponds to the choice $A^1 \equiv A^{\tilde 0}$, the
other one to $A^2 \equiv A^{\tilde 0}$.}
\label{CB_lattice}
\end{figure}

Finally, let us comment on the group structure of the transformations \eqref{e:non_abelian_trafo_vectors_4d}.
It turns out that the transformations for different choices of $A^{\tilde 0}$ all commute. We conclude that the maps \eqref{e:non_abelian_trafo_vectors_4d} generate a group $\mathbb{Z}^{\rk G}$.
It relates infinitely many classes of theories on the Coulomb branch that all lead to the same effective physics
if and only if anomalies are absent in four dimensions. This is expected when noting that \eqref{e:non_abelian_trafo_vectors_4d}
and \eqref{shift-vevs} can be associated to a higher-dimensional large gauge transformation. In fact, in an anomaly-free theory 
one therefore finds that the moduli space of the Coulomb branch VEVs $\langle\zeta^{I} \rangle$ is 
actually found to be a torus.\footnote{We like to thank Federico Bonetti for discussions on these points.}  
For a gauge group with rank two this is illustrated in \autoref{CB_lattice}.

\subsection{Abelian models} \label{sec:4d_Ab}

Let us now move to the purely Abelian theory, i.e.~we drop the fields $\hat A$ and include $\hat A^m$. In contrast to the previous section
we now face a lot more structure because there may now be a Green-Schwarz mechanism at work.
Similar to the preceding discussion we consider a basis transformation of the vectors, for which we first choose an arbitrary gauge field $A^{\tilde 0}$ out of
the $A^m$. The remaining gauge fields of $A^m$ will be denoted by $A^{\tilde m}$. The map is then defined as
\begin{align}
\label{e:abelian_trafo_vectors_4d}
 \begin{pmatrix}
  \tilde A^{0}\\[8pt]
  \tilde A^{\tilde 0}\\[8pt]
  \tilde A^{\tilde m}\\[8pt]
  \tilde A^{\alpha}
 \end{pmatrix} = 
\begin{pmatrix*}[r]
 1 & 0 & 0 & 0 \\[8pt]
 1 & 1 & 0 & 0 \\[8pt]
 0 & 0 & \delta^{\tilde m}_{\tilde n} & 0 \\[8pt]
 \frac{1}{2}b_{\tilde 0 \tilde 0}^\alpha & b_{\tilde 0 \tilde 0}^\alpha &
 b_{\tilde 0 \tilde n}^\alpha & \delta_\beta^\alpha
\end{pmatrix*}\cdot
\begin{pmatrix}
  A^{0}\\[8pt]
  A^{\tilde 0}\\[8pt]
  A^{\tilde n}\\[8pt]
  A^{\beta} \, 
 \end{pmatrix}.
\end{align}
The transformed quantities carry a tilde and
$\tilde A^{\tilde 0},\tilde A^{\tilde m}$ are again collected in $\tilde A^{m}$.
The group structure is in analogy to the last subsection $\mathbb{Z}^{n_{U(1)}}$. 
Again one expects that \eqref{e:abelian_trafo_vectors_4d} can be associated 
to a large gauge transformation in the four-dimensional theory and 
it would be nice to check this explicitly. 

This time the classical Chern-Simons terms \eqref{e:CS_class_4d} are all invariant except of
\begin{align}
 \Theta_{\alpha 0} \mapsto\tilde\Theta_{\alpha 0} = \Theta_{\alpha \tilde 0} \, .
\end{align}
Recall that in the circle reduced theory we originally had $\Theta_{\alpha 0}=0$, but $\Theta_{\alpha \tilde 0}\neq 0$
as in \eqref{e:CS_class_4d} in the presence of a Green-Schwarz mechanism.
Therefore $\tilde\Theta_{\alpha 0} \neq \Theta_{\alpha 0}$, which contradicts our requirement for the 
classical Chern-Simons coefficients to be invariant.
This issue can be cured by performing additional shifts
$\delta \tilde\Theta_{\Lambda\Sigma}$ for the Chern-Simons coefficients $\tilde\Theta_{\Lambda\Sigma}$.\footnote{
Again this can be motivated by the   M-/F-theory realization of this transformation. In \autoref{sec:F-theory} we show that
it is related to the transformation of $G_4$-flux under \eqref{e:abelian_trafo_vectors_4d}.} 
Explicitly, they read
\begin{align}\label{e:abelian_shift_4d}
 \delta \tilde\Theta_{\alpha 0} &= -\Theta_{\alpha \tilde 0}\, , \qquad 
 & \delta \tilde\Theta_{mn} &=  b^{\alpha}_{mn} \Theta_{\alpha \tilde 0}\, .
\end{align}
All other Chern-Simons coefficients remain unchanged. 
At first, these shifts appear mysterious even though they are straightforwardly motivated 
in the M-/F-theory setup. A field theory explanation exploits the fact that 
the coefficients $\Theta_{\alpha 0}$ and $\Theta_{mn}$ arise \textit{classically} if one performs a 
KK-reduction with circle fluxes for the 
axions $\hat \rho_\alpha$
\begin{subequations}
\begin{align}
 \Theta_{\alpha 0} &= \frac{1}{2} \int_{S^1}  \langle d\hat \rho_\alpha \rangle \, , \\
 \Theta_{mn}^{\rm class} &= -\frac{1}{2}b^\alpha_{mn} \int_{S^1}  \langle d\hat \rho_\alpha \rangle \, .
\end{align}
\end{subequations}
We thus realize that large gauge transformations in this case induce circle fluxes for axions. 
The shifts \eqref{e:abelian_shift_4d} undo this modification by manually switching on compensating flux in the other direction along the 
circle. As we will see in the next section, 
these shifts are not necessary in the six-dimensional setup. 

Finally, let us determine
how \eqref{e:abelian_trafo_vectors_4d} acts on the charges of the fields in the theory. For 
states $\psi_{(n)}(q)$  we find that they mix according to
 \begin{align}
\label{e:abelian_trafo_charges_4d}
 \begin{pmatrix}
  \tilde n\\[8pt]
  \tilde q_{\tilde 0}\\[8pt]
  \tilde q_{\tilde m}
 \end{pmatrix} = 
\begin{pmatrix*}[r]
 1 & 1 & 0  \\[8pt]
 0 & 1 & 0  \\[8pt]
 0 & 0 & 1 
\end{pmatrix*}\cdot
\begin{pmatrix}
  n\\[8pt]
  q_{\tilde 0}\\[8pt]
  q_{\tilde m}
 \end{pmatrix} \, .
\end{align}
As before we can apply this transformation to the calculation of one-loop Chern-Simons terms.

The one-loop
Chern-Simons coefficient $\tilde \Theta_{mn}$
can be directly evaluated in the transformed basis using \eqref{e:4d_A_regular3}
\begin{align}\label{e:4D_direct1}
 \tilde \Theta_{m n} = \sum_{q} F(q)\,q_{m} q_{n}\, \big(\tilde l_{q}+\frac{1}{2}\big)\,  \sign (\tilde m_{\rm CB}^{q}) \, ,
\end{align}
where we used $\tilde q_{m} = q_{m}$.
Mapping $\tilde \Theta_{mn}$ back to the old basis and calculating the loop there we obtain
\begin{align}\label{e:4D_indirect1}
 \tilde \Theta_{m n} &= \Theta_{ m n} 
 - b_{\tilde 0 m}^\alpha \Theta_{\alpha  n}  - b_{\tilde 0 n}^\alpha \Theta_{\alpha m} \nn \\
 &= \sum_{q} F(q)\, q_{m} q_{n}\,\big(l_{q}+\frac{1}{2})\, \sign (m_{\rm CB}^{q}\big)
 - \frac{1}{2}b_{\tilde 0 m}^\alpha \theta_{\alpha n}  - \frac{1}{2}b_{\tilde 0 n}^\alpha \theta_{\alpha m} \, .
\end{align}
We can now match \eqref{e:4D_direct1} with \eqref{e:4D_indirect1} supplemented by the shift \eqref{e:abelian_shift_4d} and use the identity
\begin{align}\label{e:charge_id}
 q_{\tilde 0} &= \big(l_{q} + \frac{1}{2}\big) \sign (m^{q}_{\rm CB}) - \big(\tilde l_q + \frac{1}{2}\big) \sign (\tilde m^q_{\rm CB}) \, ,
\end{align}
which is proven in \autoref{CB_id}. We end up with the equation
\begin{align}\label{e:Abelian_anomaly_4d}
 \sum_q F(q)\, q_{m} q_{n} q_{\tilde 0} = 
 \frac{3}{2} b_{(m n}^\alpha \theta_{\tilde 0) \alpha} \, 
\end{align}
and recognize it as a subset of the purely Abelian anomaly equations in four dimensions \eqref{4d_A_anomaly}.
Performing this matching for the other one-loop Chern-Simons coefficients $\tilde\Theta_{00}$, $\tilde\Theta_{0m}$
and picking different gauge fields for the distinguished field $A^{\tilde 0}$ in the transformation
one is able to generate the full set of Abelian anomaly equations (with repetitions).

\section{Anomalies of six-dimensional theories}\label{sec:6d_anomalies}

The techniques to obtain the anomaly equations via circle reduction developed in the previous section
work similarly in six dimensions. The generalities on the six-dimensional setup and its anomaly cancelation 
conditions are introduced in \autoref{6D_gen}. This theory is compactified on a circle in \autoref{sec:5DCSterms},
where we also recall the form of the five-dimensional Chern-Simons terms arising by integrating out massive 
modes. The Kaluza-Klein perspective on the anomaly cancelation conditions for non-Abelian and Abelian 
gauge groups is discussed in \autoref{NonAb_6Dmodels} and \autoref{Ab_6Dmodels}, respectively.
 In addition to the pure gauge anomalies we are able  
to derive the mixed gauge-gravitational anomaly equation from considering higher-curvature Chern-Simons terms.

\subsection{General setup and anomaly conditions} \label{6D_gen}

To simplify the discussion we analyze the two cases of a pure non-Abelian simple gauge group $G$ and a
pure Abelian gauge group in detail. We start first by introducing the general setup.
The gauge bosons of the simple group $G$ are denoted by $\hat A$ with components
$\hat A^\cI$, $\cI = 1,\dots ,\text{dim}\, G$ and Cartan directions $\hat A^I$, $I = 1,\dots ,\rk G$. The $n_{U(1)}$
Abelian gauge fields are denoted by $\hat A^m$ with $m =1,\dots ,n_{U(1)}$.

We introduce the chiral index $F_{\fe}(R)$ of representations $R$ of spin-$\sfrac12$ Weyl 
spinors $\hat\psi$ under $G$.
The covariant derivative is given by
\begin{align}
 \cD_\mu \hat\psi = \nabla_\mu \hat\psi - i \hat A^\cI_\mu T^R_\cI \ \hat\psi \, ,
\end{align}
where $T^R_\cI$ acts in the representation $R$ as discussed above. 
Furthermore $F_{\fe}(q)$ is the chiral index of spin-$\sfrac12$ Weyl spinors $\hat\psi$
with charges $q =(q_m)$, whose
covariant derivative reads
\begin{align}
 \cD_\mu \hat\psi = \nabla_\mu \hat\psi - i q_m \hat A^m_\mu \ \hat\psi \, .
\end{align}
In six dimensions spin-$\sfrac32$ Weyl spinors, as they e.g.~appear in supergravity multiplets, 
can be chiral with chiral index $F_{\gr}$ and therefore may be included in the anomaly analysis.
We assume that these fermions are uncharged and therefore do not contribute to anomalies 
involving the gauge fields. 
Furthermore, a six-dimensional theory can admit $T_{\mathrm{sd}}$ self-dual and $T_{\mathrm{asd}}$ 
anti-self-dual two-forms
$\hat B_\alpha$, $\alpha =1,\dots ,T_{\mathrm{sd}} + T_{\mathrm{asd}}$.
These forms carry a definite chirality and can contribute to the anomalies.
We require that the tensors are not charged,
however, they can 
couple with Green-Schwarz couplings to the gauge fields and curvature two-form
$\hat{\mathcal{R}}$ as
\begin{align}\label{e:GS_factor_6d}
 \hat{S}_{\textrm{GS}}=-\frac{1}{2} \int _{M_6}\Omega_{\alpha \beta}\hat{B}^\alpha \wedge \Big( \frac{1}{2}a^\alpha \tr\,\hat{\mathcal{R}}\wedge\hat{\mathcal{R}}
                             + 2b^\alpha_{mn}\hat{F}^m \wedge \hat{F}^n 
                             + 2\frac{b^\alpha}{\lambda (G)}\tr_f \hat{F} \wedge \hat{F}\Big) \ ,
\end{align}
where $\lambda(G)$ is defined in \autoref{app:traces}.
The coefficients $a^\alpha$, $b^\alpha_{mn}$, $b^\alpha$ are the Green-Schwarz coefficients, 
while $\Omega_{\alpha \beta}$ is constant, symmetric in its indices and its signature consists of $T_{\mathrm{sd}}$ positive signs and
$T_{\mathrm{asd}}$ negative ones. 
We stress that our considerations in the following subsections 
also apply to situations where no Green-Schwarz mechanism is applied to 
cancel anomalies. Also the coupling to gravity can be analyzed independently 
of the gauge theory analysis. Note that also
other non-chiral fields may be present. Since they neither contribute to the anomaly nor do their descendants in five dimensions
induce one-loop Chern-Simons terms, they have no effect in the following discussion.

Since we will reproduce six-dimensional gauge anomalies and mixed gravitational anomalies from loop-corrections in five dimensions,
let us display these cancelation conditions explicitly\footnote{We omit Abelian-non-Abelian anomalies, since we will not account for these in the following discussion.}
\begin{subequations}\label{e:6d_anom}
\begin{align}
 \label{e:6d_anom_1} 6  a^\alpha \frac{b^\beta}{\lambda (G)} \Omega_{\alpha\beta} &= \sum_R F_{\fe}(R) A_R  \, , \\
 \label{e:6d_anom_2}  6  a^\alpha b^\beta_{mn} \Omega_{\alpha\beta} &= \sum_{q} F_{\fe}(q) q_m q_n \, , \\
 \label{e:6d_anom_3}  0 &= \sum_R F_{\fe}(R) B_R \, , \\
 \label{e:6d_anom_4}  -3 \frac{b^\alpha}{\lambda (G)} \frac{b^\beta}{\lambda (G)} \Omega_{\alpha\beta} &= \sum_R F_{\fe}(R) C_R \, , \\
 \label{e:6d_anom_5}  -( b^\alpha_{mn} b^\beta_{pq} + b^\alpha_{mp} b^\beta_{nq} + b^\alpha_{mq} b^\beta_{np} ) \Omega_{\alpha\beta}
 &= \sum_{q} F_{\fe}(q) q_m q_n q_p q_q \, ,
\end{align}
\end{subequations}
where the constants $A_R$, $B_R$, $C_R$ are defined as
\begin{align}\label{e:trace_def}
 \tr_R \hat F^2 &= A_R \ \tr_f \hat F^2 \nn \\
 \tr_R \hat F^4 &= B_R \ \tr_f \hat F^4  + C_R \ (\tr_f \hat F^2 )^2 \, .
\end{align}

\subsection{Circle compactification and Chern-Simons terms} \label{sec:5DCSterms}

Let us now compactify the theory on a circle and push it to the Coulomb branch.
We use the conventions that six-dimensional fields have a hat, while the hat is 
dropped on all five-dimensional fields.
At the massless level
we obtain $\rk G$ five-dimensional $U(1)$ gauge fields $A^I$ from reducing $\hat A$
and $\text{dim}\, G - \rk G$ W-bosons $A^{\boldsymbol \alpha}$ labeled 
by the roots ${\boldsymbol \alpha}$ of $G$. 
In the Abelian sector we have $n_{U(1)}$ $U(1)$ gauge fields $A^m$ from the reduction of the $\hat A^m$.
To the $\text{dim}\, G+n_{U(1)}$ gauge fields one finds associated Wilson line 
scalars $\zeta^\cI$, $\zeta^m$. The expansions of 
the six-dimensional gauge fields are analogous to \eqref{red-AI}. The scalars $\zeta^\cI$, $\zeta^m$ 
parametrize the five-dimensional Coulomb branch
\beq
     \langle \zeta^I \rangle \neq 0 \ , \qquad \langle \zeta^{\boldsymbol \alpha} \rangle = 0 \ , \qquad  \langle \zeta^m \rangle \neq 0 \ .
\eeq
These expectation values break the gauge group as $G \times U(1)^{n_{U(1)}} \rightarrow U(1)^{\rk G} \times U(1)^{n_{U(1)}}$ 
while rendering the W-bosons massive. The six-dimensional metric also contains the KK-vector $A^0$, the 
radius $r$ and a five-dimensional metric $g_{\mu \nu}$ in analogy to \eqref{e:metric_reduction}.
For simplicity we assume that $g_{\mu \nu}$ is Minkowskian.
Furthermore, we find $T_{\mathrm{sd}}+T_{\mathrm{asd}}$ Abelian vectors $A^\alpha$ from reducing and dualizing $\hat B_\alpha$.
The details on this reduction can be found, for example, in refs.~\cite{Bonetti:2011mw,Grimm:2013oga}.

Considering the theory on the Coulomb branch, the charged fields and the W-bosons generically become massive. 
In addition, higher KK-modes admit a KK-mass contribution proportional to their level $n$. In summary, and in complete analogy to \eqref{e:KK-masses},
the fields in the five-dimensional theory acquire masses
\begin{align}
 m = \begin{cases}
      m^w_{\rm CB} + n \,m_{\rm KK} = w_I   \langle \zeta^I \rangle+ \frac{n}{\langle r \rangle}, 
      \qquad &\textrm{non-Abelian gauge group}\, , \\
      m^q_{\rm CB} + n \,m_{\rm KK} = q_m \langle \zeta^m \rangle + \frac{n}{\langle r\rangle},  &\textrm{Abelian gauge group}  \, .
     \end{cases}
\end{align}

In determining the effective theory for the massless modes one needs to systematically integrate out all massive 
states. As in three dimensions we focus on the Chern-Simons terms. These terms are topological in nature 
and can be corrected at one-loop level by all massive states. Five-dimensional Chern-Simons terms for 
$U(1)$ gauge fields take the general form
\begin{subequations}\label{e:def_CS_6d}
\begin{align}
 S^{\textrm{gauge}}_{\textrm{CS}} = - \frac{1}{12} \int k_{\Lambda\Sigma\Theta} A^{\Lambda} \wedge F^{\Sigma} \wedge F^{\Theta} \\
 S^{\textrm{grav}}_{\textrm{CS}} = - \frac{1}{4} \int k_{\Lambda} A^{\Lambda} \wedge \tr ( \mathcal{R} \wedge \mathcal{R} ) \, ,
\end{align}
\end{subequations}
where $k_{\Lambda\Sigma\Theta} $ and $k_{\Lambda} $ are constants and  $\mathcal{R}$
is the five-dimensional curvature two-form. 
The collective index is $\Lambda = (0,I,\alpha)$ or $\Lambda = (0,m,\alpha)$, respectively.
The classical Chern-Simons coefficients in the circle reduced theory are found to be
\begin{align}\label{e:CS_cl_6d}
 &k_{\alpha\beta\gamma} =0 \ , &&k_{ 0\alpha\beta} = \Omega_{\alpha\beta} \ , &&k_{I\alpha\beta} =0 \ , \nn \\
 &k_{m\alpha\beta} =0 \ , && k_{00\alpha} =0 \ , && k_{IJ\alpha} = -\Omega_{\alpha\beta} b^\beta \cC_{IJ} \ , \nn \\
 &k_{mn\alpha} = -\Omega_{\alpha\beta} b^\beta_{mn} \ , &&k_{0I\alpha} =0 \ , &&k_{0m\alpha} =0 \ ,\nn \\ 
 &k_{\alpha} = -12\ \Omega_{\alpha\beta} a^\beta \ .
\end{align}
with $\cC_{IJ}$ the coroot intersection matrix defined in \eqref{e:def_coroot_int_mat}.
Note that since we restrict either to pure non-Abelian or pure Abelian gauge groups, we do not display Chern-Simons coefficients involving both
types of indices $m,n$ and $I,J$.

In addition to the classical Chern-Simons terms \eqref{e:CS_cl_6d} the 
effective theory can admit one-loop induced Chern-Simons terms from massive charged two-forms, spin-$\sfrac12$ and spin-$\sfrac32$ fermions
\cite{Witten:1996qb,Bonetti:2012fn,Bonetti:2013ela}. Their contributions are given by
\begin{align}
\label{e:6d_single_field_CS_1} k_{\Lambda\Sigma\Theta}^{\textrm{loop}} = c_{AFF}\, q_\Lambda q_\Sigma q_\Theta \, \sign (m) \, , \\
\label{e:6d_single_field_CS_2} k_{\Lambda}^{\textrm{loop}} = c_{A\cR\cR}\, q_\Lambda \, \sign (m) \, ,
\end{align}
where the $c_{AFF},c_{A\cR\cR}$ are given in \autoref{t:CS_correct}.
\begin{table}
\begin{center}
\begin{tabular}{llll}
\hline\hline
 & spin-$\sfrac{1}{2}$ fermion & self-dual tensor & spin-$\sfrac{3}{2}$ fermion\\
 \hline
$c_{AFF}$ & $1/2$ & $-2$ & $5/2$\\
$c_{A\cR\cR}$ & $-1$ & $-8$ & $19$\\
\hline\hline
\end{tabular}
\end{center}
\caption{One-loop Chern-Simons term factors.}
\label{t:CS_correct}
\end{table}
The quantity $q_\Lambda$ is the charge under $A^\Lambda$ and $\sign (m)$ depends on the representation of the massive little group
$SO(4) \cong SU(2) \times SU(2)$ in the following way
\begin{align}
 \sign (m) = \begin{cases}
              +1 \quad \textrm{for } (\frac{1}{2},0),(1,0),(1,\frac{1}{2})\, , \\
              -1 \quad \textrm{for } (0,\frac{1}{2}),(0,1),(\frac{1}{2},1) \, ,
             \end{cases}
\end{align}
where we labeled representations of $SU(2) \times SU(2)$ by their spins.
In order to explicitly compute these terms we use the convention that the charge under the KK-vector 
is given as in \eqref{e:KK_charge_convention}.
As in the preceding section we exploit zeta function regularization in order to treat the infinite KK-tower.
Along the lines of \cite{Grimm:2013oga} the relevant one-loop
Chern-Simons coefficients of the circle reduced theory for the pure non-Abelian
and pure Abelian theory respectively are evaluated to be
\begin{subequations}\label{e:6d_regular}
\begin{align}
 \label{e:6d_na_regular4}k_{IJK} &= \frac{1}{2}\sum_R F_{\fe}(R) \sum_{w \in R} \ (2l_w +1 ) \ w_I w_J w_K \ \sign  (m^w_{\rm CB}) \, , \\
 \label{e:6d_na_regular6}k_{I} &= -\sum_R F_{\fe}(R) \sum_{w \in R} \ (2l_w +1 ) \ w_I \ \sign  (m^w_{\rm CB}) \, , \\
 \label{e:6d_a_regular4}k_{mnp} &= \frac{1}{2}\sum_{q}  F_{\fe}(q) \ (2l_q +1 ) \ q_m q_n q_p \ \sign  (m^q_{\rm CB}) \, , \\
 \label{e:6d_a_regular6}k_{m} &= -\sum_{q}  F_{\fe}(q) \ (2l_q +1 ) \ q_m \ \sign  (m^q_{\rm CB}) \, ,
\end{align}
\end{subequations}
where $l_w$, $l_q$ are defined in \eqref{e:def_lq}.
The remaining one-loop Chern-Simons terms are listed for completeness in \autoref{app:loops}.
Note that the classical Chern-Simons coefficients do not receive any corrections.
We proceed by establishing the precise correspondence of these one-loop Chern-Simons coefficients \eqref{e:6d_regular}
and the anomaly conditions \eqref{e:6d_anom}.

\subsection{Non-Abelian models} \label{NonAb_6Dmodels}

We start by considering the purely non-Abelian simple gauge group $G$ without Abelian factors
and apply a basis transformation on the vectors in the circle reduced theory on the Coulomb branch.
In analogy to the discussion of \autoref{sec:4d_anomalies}, we first pick an arbitrary vector $A^{\tilde 0}$ out of the Cartan 
elements $A^{I}$. The remaining Cartan vectors are denoted by $A^{\tilde I}$, while the remaining 
$\text{dim}\,G-1$ vectors of the $A^\cI$ are denoted by $A^{\tilde \cI}$.
The transformation then takes the form 
\begin{align}
 \label{e:non_abelian_trafo_vectors_6d}
 \begin{pmatrix}
  \tilde A^{0}\\[8pt]
  \tilde A^{\tilde 0}\\[8pt]
  \tilde A^{\tilde \cI}\\[8pt]
  \tilde A^{\alpha}
 \end{pmatrix} = 
\begin{pmatrix*}[r]
 1 & 0 & 0 & 0 \\[8pt]
 1 & 1 & 0 & 0 \\[8pt]
 0 & 0 & \delta^{\tilde \cI}_{\tilde \cJ} & 0 \\[8pt]
 \frac{1}{2}b^\alpha \cC_{\tilde 0 \tilde 0} & b^\alpha \cC_{\tilde 0 \tilde 0} &
 b^\alpha \cC_{\tilde 0 \tilde \cJ} & \delta_\beta^\alpha
\end{pmatrix*}\cdot
\begin{pmatrix}
  A^{0}\\[8pt]
  A^{\tilde 0}\\[8pt]
  A^{\tilde \cJ}\\[8pt]
  A^{\beta} \, 
 \end{pmatrix},
\end{align}
where $\cC_{IJ}$ is the coroot intersection matrix \eqref{e:def_coroot_int_mat} of $G$ and $\cC_{\tilde \cI \tilde \cJ}$ vanishes for all 
non-Cartan directions.
We denote quantities in the transformed basis by a tilde and again collect $\tilde A^{\tilde 0}$, $\tilde A^{\tilde I}$
in $\tilde A^{I}$.
The group structure is simply $\mathbb{Z}^{\rk G}$.
One can check that the classical Chern-Simons terms \eqref{e:CS_cl_6d} remain invariant under this transformation
and the weights $w$ of some representation $R$ transform as
in \eqref{e:non_abelian_trafo_charges_4d}.

Let us next apply \eqref{e:non_abelian_trafo_vectors_6d} to the one-loop Chern-Simons coefficients. Concretely, 
we first evaluate
$\tilde k_{IJK}$ by performing the loop-computation directly in the transformed basis 
making use of \eqref{e:6d_na_regular4} to find
\begin{align}\label{e:direct_calc_6d_non_Ab}
 \tilde k_{IJK} = \frac{1}{2}\sum_R F_{\fe}(R) \sum_{w \in R}  \ (2\tilde l_w +1 ) \ w_{I} w_{J}
 w_{K} \ \sign  (\tilde m^w_{\rm CB}) \, .
\end{align}
Then we can also transform $\tilde k_{IJK}$ back to the old basis and perform the loop-computation there
\begin{align}\label{e:indirect_calc_6d_non_Ab}
 \tilde k_{IJK} =\ & k_{IJK} - k_{I J \alpha} b^\alpha \cC_{\tilde 0 K}
  - k_{I K \alpha} b^\alpha \cC_{\tilde 0 J} - k_{J  K \alpha} b^\alpha \cC_{\tilde 0 I} \nn \\
   =\ & \frac{1}{2}\sum_{R}  F_{\fe}(R) \sum_{w \in R} \ (2 l_w +1 ) \ w_{I} w_{J}  w_{K} \ \sign  ( m^w_{\rm CB}) \nn \\
  & + b_{\alpha} b^\alpha  \cC_{I J} \cC_{\tilde 0 K}
  + b_{\alpha}  b^\alpha  \cC_{I K} \cC_{\tilde 0 J} + b_{\alpha} b^\alpha  \cC_{J K} \cC_{\tilde 0 I} \, .
\end{align}
Matching \eqref{e:direct_calc_6d_non_Ab} with \eqref{e:indirect_calc_6d_non_Ab} we obtain using \eqref{e:weight_id}
\begin{align}\label{e:6D_CS_match}
   \sum_{R}  F_{\fe}(R) \sum_{w \in R} \ w_{I} w_{J}  w_{K} w_{\tilde 0} = - b_{\alpha} b^\alpha \cC_{I J}
  \cC_{\tilde 0 K}
 -  b_{\alpha} b^\alpha  \cC_{I K} \cC_{\tilde 0 J} - b_{\alpha} b^\alpha  \cC_{J K} \cC_{\tilde 0 I} \, .
\end{align}
In \autoref{app:traces} it is shown that this equation is equivalent to both pure non-Abelian gauge anomaly cancelation
conditions \eqref{e:6d_anom_3}, \eqref{e:6d_anom_4}.
More precisely, we show in \autoref{app:traces} that it is possible to rewrite non-Abelian anomaly cancelation conditions in a
convenient fashion.
While in the usual presentation these conditions are written in terms of the factors $A_R$, $B_R$, $C_R$, which appear in the reduction of traces
\eqref{e:trace_def}, we managed to reformulate them in terms of sums over weights, not involving any of the factors $A_R$, $B_R$, $C_R$.
Applying this procedure to the remaining gauge one-loop Chern-Simons coefficients
again yields the anomaly conditions \eqref{e:6d_anom_3}, \eqref{e:6d_anom_4}.

Interestingly we can also reproduce the mixed non-Abelian-gravitational anomaly \eqref{e:6d_anom_1} by investigating
the behavior of the gravitational Chern-Simons coefficient $\tilde k_{I}$. On the one hand, by directly evaluating the loop we obtain
using \eqref{e:6d_na_regular6}
\begin{align}\label{e:grav_direct}
 \tilde k_{I} = -\sum_R F_{\fe}(R) \sum_{w \in R} \ (2\tilde l_w +1 ) \ w_{I} \ \sign  (\tilde m^w_{\rm CB})\, .
\end{align}
On the other hand, in the old basis we find
\begin{align}\label{e:grav_indirect}
  \tilde k_{I} = k_{I} - b^\alpha \cC_{\tilde 0 I} k_\alpha = 
  -\sum_R F_{\fe}(R) \sum_{w \in R} \ (2l_w +1 ) \ w_{I} \ \sign  (m^w_{\rm CB}) + 12 b^\alpha a_\alpha  \cC_{\tilde 0 I} \, .
\end{align}
Using \eqref{e:weight_id} the matching of \eqref{e:grav_direct} and \eqref{e:grav_indirect} yields the condition
\begin{align}\label{e:nA_anomaly_CS}
  \sum_R F_{\fe}(R) \sum_{w \in R} \ w_{I} w_{\tilde 0} = 6 b^\alpha a_\alpha  \cC_{\tilde 0 I} \, .
\end{align}
It is shown in \cite{Grimm:2013oga} that the following identity holds true
\begin{align}
 \sum_{w \in R} \ w_{I} w_{J} = A_R \,\lambda (G) \,\cC_{IJ} \quad \forall I,J \, .
\end{align}
Thus we can conclude that \eqref{e:nA_anomaly_CS} coincides with the mixed non-Abelian-gravitational anomaly \eqref{e:6d_anom_1}.
Applying these steps to $k_0$ again
yields the mixed anomaly.

\subsection{Abelian models} \label{Ab_6Dmodels}

Let us finally investigate a purely Abelian theory.
In complete analogy we pick one arbitrary vector $A^{\tilde 0}$ out of the $A^{m}$ and denote the remaining ones
by $A^{\tilde m}$. We define the transformation
\begin{align}
\label{e:abelian_trafo_vectors_6d}
 \begin{pmatrix}
  \tilde A^{0}\\[8pt]
  \tilde A^{\tilde 0}\\[8pt]
  \tilde A^{\tilde m}\\[8pt]
  \tilde A^{\alpha}
 \end{pmatrix} = 
\begin{pmatrix*}[r]
 1 & 0 & 0 & 0 \\[8pt]
 1 & 1 & 0 & 0 \\[8pt]
 0 & 0 & \delta^{\tilde m}_{\tilde n} & 0 \\[8pt]
 \frac{1}{2}b_{\tilde 0 \tilde 0}^\alpha & b_{\tilde 0 \tilde 0}^\alpha &
 b_{\tilde 0 \tilde n}^\alpha & \delta_\beta^\alpha
\end{pmatrix*}\cdot
\begin{pmatrix}
  A^{0}\\[8pt]
  A^{\tilde 0}\\[8pt]
  A^{\tilde n}\\[8pt]
  A^{\beta} \, ,
 \end{pmatrix},
\end{align}
which has precisely the same form as \eqref{e:abelian_trafo_vectors_4d} and is therefore isomorphic to $\mathbb{Z}^{n_{U(1)}}$.
The quantities in the transformed basis carry a tilde.
However, in contrast to the four-dimensional setup, this transformation leaves all classical Chern-Simons terms \eqref{e:CS_cl_6d}
invariant and there is no need for an additional shift like \eqref{e:abelian_shift_4d}.\footnote{The reason for this can be understand via the
M-/F-theory realization, since this time there is no $G_4$-flux that needs to be transformed.} Finally the charges transform
as in \eqref{e:abelian_trafo_charges_4d}.

We now show that the transformation of one-loop Chern-Simons terms under \eqref{e:abelian_trafo_vectors_6d} yields the pure Abelian
anomaly equations, as well as the mixed Abelian-gravitational anomaly.
First consider $\tilde k_{mnp}$ and evaluate the loop directly using \eqref{e:6d_a_regular4} as
\begin{align}\label{e:direct_calc_6d_Ab}
 \tilde k_{mnp} = \frac{1}{2}\sum_{q}  F_{\fe}(q) \ (2\tilde l_q +1 ) \ q_{m} q_{n}
 q_{p} \ \sign  (\tilde m^q_{\rm CB}) \, .
\end{align}
On the other hand we can transform $\tilde k_{mnp}$ back to the old basis and do the loop there
\begin{align}\label{e:indirect_calc_6d_Ab}
 \tilde k_{mnp}&=   k_{mnp} - k_{mn \alpha} b^\alpha_{\tilde 0 p}
  - k_{m p \alpha} b^\alpha_{\tilde 0 n} - k_{n p \alpha} b^\alpha_{\tilde 0 m} \nn \\
  &=  \frac{1}{2}\sum_{q}  F_{\fe}(q) \ (2 l_q +1 ) \ q_{m} q_{n}  q_{p} \ \sign  ( m^q_{\rm CB}) 
  + ( b_{mn}^\alpha b^\beta_{\tilde 0 p}
  + b_{m p}^\alpha b^\beta_{\tilde 0 n} + b_{n p}^\alpha b^\beta_{\tilde 0 m})\Omega_{\alpha\beta} \, .
\end{align}
Let us now match \eqref{e:direct_calc_6d_Ab} and \eqref{e:indirect_calc_6d_Ab} and use the identity \eqref{e:charge_id} to obtain the equation
\begin{align}
  \sum_{q}  F_{\fe}(q) \ q_{m} q_{n}  q_{p} q_{\tilde 0} = - ( b_{mn}^\alpha b^\beta_{\tilde 0 p}
 +  b_{mp}^\alpha b^\beta_{\tilde 0 n} + b_{np}^\alpha b^\beta_{\tilde 0 m} ) \Omega_{\alpha\beta} \, ,
\end{align}
which is the pure Abelian anomaly condition \eqref{e:6d_anom_5}. Treating also the remaining gauge one-loop Chern-Simons
coefficients in this way and making different choices for the vector
$A^{\tilde 0}$ in the transformation \eqref{e:abelian_trafo_vectors_6d} we obtain all pure Abelian anomaly conditions (with repetitions).

Lastly we can perform the loop-computation for $\tilde k_{m}$ \eqref{e:6d_a_regular6}
\begin{align}
 \tilde k_{m} = -\sum_{q}  F_{\fe}(q) \ (2\tilde l_q +1 ) \ q_m \ \sign  (\tilde m^q_{\rm CB})
\end{align}
or transform it back again
\begin{align}
 \tilde k_{m} = k_{m} - b^\alpha_{\tilde 0 m} k_\alpha = 
  -\sum_q F_{\fe}(q) \ (2l_q +1 ) \ q_{m} \ \sign  (m^q_{\rm CB}) + 12 b^\alpha_{\tilde 0 m} a_\alpha \, .
\end{align}
Using \eqref{e:charge_id} this results in the matching
\begin{align}
  \sum_q F_{\fe}(q) \ q_{m} q_{\tilde 0} = 6 b^\alpha_{\tilde 0 m} a^\beta \Omega_{\alpha\beta} \, ,
\end{align}
which coincides with the mixed Abelian-gravitational anomaly equation \eqref{e:6d_anom_2}.

\section{Anomalies on the circle in the M-/F-theory duality}\label{sec:F-theory}

In this section we provide the motivation for the transformations that led
to the appearance of the higher-dimensional anomaly equations in the Kaluza-Klein theories on the circle.
More precisely, we consider F-theory compactifications on elliptically fibered Calabi-Yau manifolds and investigate
aspects of the effective action exploiting the duality to M-theory. 
After reviewing some properties of F-theory compactifications on elliptically fibered Calabi-Yau manifolds in \autoref{CYgeometry}, 
we show that the transformations considered in \autoref{sec:4d_anomalies} and \autoref{sec:6d_anomalies} correspond to geometric symmetries. 

The analysis for Abelian gauge symmetries is provided in \autoref{zero-section-section}. We show
that one is free to choose an arbitrary section of the elliptic fibration as
the \textit{`zero-section'} and that different choices for zero-sections are related by transformations
similar to \eqref{e:abelian_trafo_vectors_4d}, \eqref{e:abelian_trafo_vectors_6d}.
Thus for Calabi-Yau fourfolds and threefolds the invariance of F-theory compactifications under zero-section changes proves the absence of Abelian anomalies in the effective action.

We transfer this observation to the non-Abelian case in \autoref{zero-node-section}.
More precisely, we start with
transformations like \eqref{e:non_abelian_trafo_vectors_4d}, \eqref{e:non_abelian_trafo_vectors_6d}
and find retroactively their geometric manifestation. It turns out that these transformations map
between different choices of what we call the  \textit{`zero-node'}. When one resolves the singularities 
over a seven-brane divisor in the base
the elliptic fiber splits into $\rk G + 1$ irreducible components. One arbitrary component is then to be chosen as the zero-node.
However, we claim that F-theory compactifications on elliptically fibered Calabi-Yau manifolds are 
invariant under zero-node changes which then implies the absence of non-Abelian gauge anomalies.

\subsection{F-theory on elliptically fibered Calabi-Yau manifolds} \label{CYgeometry}

Let us review some basic facts about F-theory compactifications and their effective actions.
F-theory compactified on elliptically fibered Calabi-Yau fourfolds yields $\cN =1$ supergravity theories 
in four dimensions, while compactifications
on elliptically fibered Calabi-Yau threefolds yield $\cN = (1,0)$ supergravity theories in six dimensions.
In both dimensions non-Abelian gauge groups can arise from singularities of the elliptic fibration, whereas
Abelian gauge groups are induced if the fibration admits at least two sections.

In order to derive
the effective action one can employ the duality between F-theory and M-theory. More concretely,
one first reduces a general four- or six-dimensional supergravity theory on a circle, pushes the lower-dimensional 
theory onto the Coulomb branch, and matches it with the dual M-theory description. 
M-theory can be accessed via eleven-dimensional supergravity if one uses the same Calabi-Yau space, but 
with all singularities being resolved \cite{Vafa:1996xn,Morrison:1996na,Morrison:1996pp,Ferrara:1996wv,Denef:2008wq,Grimm:2010ks,Bonetti:2011mw,Grimm:2013oga}. 
The derivation of the F-theory effective actions requires to match the supergravity data with 
geometric quantities using the three- or five-dimensional effective theories. 
Importantly, the naive matching procedure fails if one only restricts the considerations 
to the classical effective theory. In particular, the classical Chern-Simons terms 
in the circle reduced supergravity theory will not directly match with the M-theory side. 
In accord with the discussion of the previous sections one recalls that one-loop 
corrections to the Chern-Simons terms are induced by massive modes. These 
have to be taken into account to match the M-theory and F-theory 
reduction \cite{Grimm:2011fx,Bonetti:2011mw,Cvetic:2012xn,Grimm:2013oga,Cvetic:2013uta,
Anderson:2014yva}.
Since the one-loop corrections to the Chern-Simons terms on the 
F-theory side are sensitive to the spectrum, the matching with the M-theory reduction 
reveals information about the spectrum in the F-theory effective theory 
in terms of geometric quantities of the resolved Calabi-Yau space and the background flux. 

Let us further focus on Chern-Simons terms and study
the matching procedure more precisely.
First we have to introduce some geometric properties of the M-theory compactification. 
The resolved Calabi-Yau manifold is denoted by $\hat Y$ and arises from an elliptic fibration over some base 
$B$. The projection to the base is denoted by $\pi: \hat Y \rightarrow B$. 
The linearly independent sections of the elliptic fibration are denoted by $\sigma_0$, $\sigma_m$, where we singled out
one arbitrary section $\sigma_0$ as the so-called zero-section.
We assume that there is always at least one section.
Next, there may be a divisor in the base $B$ of the elliptic fibration 
over which the resolved singular fiber splits into a number of irreducible components
which intersect as the affine Dynkin diagram of the gauge algebra. Fibering
the latter over this divisor in $B$ we obtain divisors of the whole fibration, which we denote by
$\Sigma_0$, $\Sigma_I$. We again singled out an arbitrary component 
$\Sigma_0$ as what we call the zero-node.\footnote{This is reminiscent
of the affine node in the F-theory literature, the node which is intersected by the zero-section.
The zero-node may be considered as a generalization of this concept, since its choice is arbitrary and it is in particular
not immediately related to the affine node of extended Dynkin diagrams.}

We can now define a convenient basis of divisors
$D_\Lambda = (D_0 , D_I , D_m , D_\alpha)$ in the elliptically fibered, resolved Calabi-Yau manifold $\hat Y$ in the following way:
\begin{itemize}
 \item The divisor $D_0$ is obtained from $\sigma_0$ supplemented by the shift \eqref{e:base_shifts}
 or the generalized shift \eqref{e:base_shifts_nA}, respectively.
 Expanding the M-theory three-form $C_3$
 along the corresponding two-form yields a vector that is identified with the KK-vector in the dual circle reduced F-theory setting.
 
 \item The exceptional divisors $D_I$ are related to the fiber components $\Sigma_I$ via \eqref{e:Shioda_shifts_nA}.
 They correspond to the Cartan generators in the dual F-theory setting
 such that $I=1,\dots , \rk G$.

 \item The $U(1)$ divisors $D_m$ descend from the sections $\sigma_m$ using the Shioda map \eqref{e:Shioda_map}.
 They correspond to $U(1)$ gauge symmetries in the F-theory setting implying that $m=1,\dots , n_{U(1)}$.
 
 \item The vertical divisors $D_\alpha$ are obtained as $\pi^{-1}(D^b_\alpha)$ from divisors $D^b_\alpha$ 
 of the base $B$. For each homologically independent divisor $D^b_\alpha$ in $B$ one finds an axion in four-dimensional F-theory compactifications and an (anti-) self-dual tensor in six dimensions, respectively.
\end{itemize}
For Calabi-Yau fourfolds we introduce vertical four-cycles $\cC^\alpha$ that are obtained as $\pi^{-1}(\cC_b^\alpha)$
from curves $\cC_b^\alpha$ in the base intersecting the $D^b_\alpha$ as
\begin{align}\label{e:def_metric}
 \tensor{\eta}{_\alpha^\beta} = D^b_\alpha \cdot \cC^\beta_b
\end{align}
with $\tensor{\eta}{_\alpha^\beta}$ a full-rank matrix.

It turns out that the Chern-Simons coefficients \eqref{e:def_CS_4d}, \eqref{e:def_CS_6d} in the M-theory compactification are given by
the intersections \footnote{In the following Poincar\'e duality is always understood implicitly.}
\cite{Ferrara:1996wv,Antoniadis:1997eg,Haack:2001jz}
\begin{align}
\label{e:M-theory_CS} \Theta_{\Lambda\Sigma} &= -\frac{1}{4} D_\Lambda \cdot D_\Sigma \cdot G_4  \qquad &&\textrm{three dimensions}, \\
 k_{\Lambda\Sigma\Theta} &= D_\Lambda \cdot D_\Sigma \cdot D_\Theta &&\textrm{five dimensions} , \\
 k_{\Lambda} &= D_\Lambda \cdot c_2 &&\textrm{five dimensions} ,
\end{align}
where $G_4$ denotes the flux of the four-form field strength and $c_2$ is the second Chern class of the Calabi-Yau threefold $\hat Y$.
It is now possible to match these quantities with their counterparts in the circle reduced supergravity theory on the Coulomb branch.
In the following we use the field theory notation as introduced in \autoref{sec:4d_anomalies} and \autoref{sec:6d_anomalies}.
The matching of intersection numbers with classical terms in the circle reduced theory gives for the fourfold \cite{Grimm:2010ks,Grimm:2012yq,Cvetic:2012xn}
\begin{align}\label{e:cl_matching_3d}
D_I \cdot D_J \cdot \cC^\alpha  &= - \cC_{IJ}\ b^\beta \tensor{\eta}{_\beta^\alpha} \, , & D_m \cdot D_n \cdot \cC^\alpha  &= - b_{mn}^\beta
\tensor{\eta}{_\beta^\alpha} \, , \nn \\
 && D_\alpha \cdot D_m \cdot G_4 &= -2 \theta_{\alpha m} \, ,
\end{align}
where $\theta_{\alpha m}$ are the axion gaugings defined in \eqref{e:axion_gauging},
$b_{mn}^\alpha$, $b^\alpha$ are the Green-Schwarz coefficients \eqref{e:4d_GS}, and $\tensor{\eta}{_\beta^\alpha}$
is the matrix defined in \eqref{e:def_metric}.
For the threefold the classical Chern-Simons matching yields \cite{Bonetti:2011mw,Grimm:2012yq,Grimm:2013oga}
\begin{align}\label{e:cl_matching_5d}
 D_0 \cdot D_\alpha \cdot D_\beta &= \Omega_{\alpha\beta} \, , & D_\alpha \cdot c_2 &= - 12 a^\beta \Omega_{\alpha\beta} \, , \\
  D_I \cdot D_J \cdot D_\alpha &= -\cC_{IJ}\ b^\beta \Omega_{\alpha\beta} \, , 
 & D_m \cdot D_n \cdot D_\alpha &= - b^\beta_{mn} \Omega_{\alpha\beta} \, , \nn 
\end{align}
where $a^\alpha$, $b^\alpha$, $b^\alpha_{mn}$ are the Green-Schwarz coefficients defined in \eqref{e:GS_factor_6d}.\footnote{Anomaly cancelation in six-dimensional F-theory compactifications has been studied in \cite{Taylor:2011wt,Sadov:1996zm,Kumar:2009ac,Park:2011wv,Park:2011ji,Grimm:2012yq}.}

As already stressed, in the circle reduced theory one-loop corrections to the Chern-Simons terms 
have to be taken into account in order to perform a complete matching. 
For simplicity let us consider only the following matching of one-loop Chern-Simons terms in three dimensions
\begin{align}
 D_I \cdot D_J \cdot G_4 =& -4  \sum_{R}  C(R) \sum_{w \in R} \ (l_w +\frac{1}{2} ) \ w_I w_J \ \sign  (m^w_{\rm CB})  \\
 &-4  \sum_{\boldsymbol{\alpha} \in \mathrm{Adj}} \ (l_{\boldsymbol{\alpha}} +\frac{1}{2} ) \ \boldsymbol{\alpha}_I
 \boldsymbol{\alpha}_J \ \sign  (m^{\boldsymbol{\alpha}}_{\rm CB}) \, , \nn \\
 D_m \cdot D_n \cdot G_4 =& -4  \sum_{q}  C(q) \ (l_q +\frac{1}{2} ) \ q_m q_n \ \sign  (m^q_{\rm CB}) \, ,
\end{align}
and in five dimensions
\begin{align} 
 D_I \cdot D_J \cdot D_K =& - \frac{1}{2}\sum_{R}  H(R) \sum_{w\in R} \ (2l_w +1 ) \ w_I w_J w_K \ \sign  (m^w_{\rm CB})  \\
 & + \frac{1}{2} \sum_{\boldsymbol{\alpha} \in \mathrm{Adj}} \ (2l_{\boldsymbol{\alpha}} +1 ) \ \boldsymbol{\alpha}_I
 \boldsymbol{\alpha}_J \boldsymbol{\alpha}_K \ \sign  (m^{\boldsymbol{\alpha}}_{\rm CB}) \, ,\nn \\
 D_m \cdot D_n \cdot D_p =& - \frac{1}{2}\sum_{q}  H(q) \ (2l_q +1 ) \ q_m q_n q_p \ \sign  (m^q_{\rm CB}) \, , \\
 D_I \cdot c_2 =& \sum_{R}  H(R) \sum_{w \in R} \ (2l_w +1 ) \ w_I \ \sign  (m^w_{\rm CB})
 + \sum_{\boldsymbol{\alpha} \in \mathrm{Adj}} \ (2l_{\boldsymbol{\alpha}} +1 ) \ \boldsymbol{\alpha}_I \
 \sign  (m^{\boldsymbol{\alpha}}_{\rm CB}) \, ,
 \end{align}
 \begin{align}
 D_m \cdot c_2 =& \sum_{q}  H(q) \ (2l_q +1 ) \ q_m \ \sign  (m^q_{\rm CB}) \, . \hspace*{3cm}
\end{align}
Here $C(R)$/$H(R)$ denote the number of chiral multiplets and hypermultiplets that 
transform in some representation $R$ under the non-Abelian gauge group $G$. Accordingly, 
$C(q)$/$H(q)$ counts the chiral multiplets and hypermultiplets with charge $q$ under the $U(1)$ gauge group factors.
Of course, there are further equations appearing in the matching of one-loop terms with the intersection numbers. 
For simplicity we will not display all matchings in the following discussion, but rather include 
some additional comments on these later.

\subsection{Abelian anomalies from zero-section changes} \label{zero-section-section}

In this subsection we show how transformations similar to \eqref{e:abelian_trafo_vectors_4d}, \eqref{e:abelian_trafo_vectors_6d} 
arise in the duality between M-theory and F-theory through different choices of the zero-section.
For simplicity we consider only the pure Abelian case in the following. F-theory compactifications 
with both Abelian and non-Abelian groups have recently been studied intensively in \cite{Grimm:2010ez,Morrison:2012ei,Braun:2013yti,Cvetic:2013nia,Borchmann:2013jwa,Grimm:2013oga,Braun:2013nqa,Borchmann:2013hta,Cvetic:2013qsa,Kuntzler:2014ila,Klevers:2014bqa,Braun:2014qka,Lawrie:2014uya}

As already announced, for the matching of the dual theories to work, one has to shift the zero-section $\sigma_0$ 
in the definition of the base divisor.
The correct way to do this is given by \cite{Grimm:2011sk,Park:2011ji,Cvetic:2012xn,Grimm:2013oga}
\begin{subequations}\label{e:base_shifts}
\begin{align}
 \label{e:base_shift_3d} D_0 &= \sigma_0 - \frac{1}{2} ( \sigma_0 \cdot \sigma_0 \cdot \cC^\alpha )
 \tensor{\eta}{^{-1}_\alpha^\beta} D_\beta \qquad &&\textrm{three dimensions}, \\
 \label{e:base_shift_5d} D_0 &= \sigma_0 - \frac{1}{2} ( \sigma_0 \cdot \sigma_0 \cdot D^\alpha )D_\alpha \qquad &&\textrm{five dimensions}. 
\end{align}
\end{subequations}
Recall that expanding the M-theory three-form along $D_0$ gives the dual to the KK-vector.
Furthermore, the remaining sections $\sigma_m$ have to be shifted using the Shioda map in order to get the
appropriate $U(1)$ divisors \cite{shioda1989mordell,shioda1990mordell}.
In absence of non-Abelian gauge symmetries the maps read
\begin{subequations}\label{e:Shioda_map}
\begin{align}
\label{e:Shioda_map_3d} D_m &= \sigma_m - D_0 - ( \sigma_m \cdot D_0 \cdot \cC^\alpha  )
 \tensor{\eta}{^{-1}_\alpha^\beta} D_\beta \qquad &&\textrm{three dimensions}, \\
\label{e:Shioda_map_5d} D_m &= \sigma_m - D_0 - ( \sigma_m \cdot D_0 \cdot D^\alpha  ) D_\alpha \qquad &&\textrm{five dimensions}.
\end{align}
\end{subequations}
It is now interesting to ask how these divisors transform if one chooses a different zero-section. Concretely
fix one $\sigma_{\tilde 0}$ out of the $\sigma_{m}$. The remaining $U(1)$ sections will be denoted by $\sigma_{\tilde m}$.
Let us then take the original zero-section $\sigma_0$ as an ordinary $U(1)$ section and select $\sigma_{\tilde 0}$ as 
the new zero-section. The corresponding transformation of the divisors, dictated 
by \eqref{e:base_shifts}, \eqref{e:Shioda_map} is found to be
\begin{align}\label{e:zero_section_change}
 \begin{pmatrix}
  \tilde D_{0}\\[8pt]
  \tilde D_{\tilde 0}\\[8pt]
  \tilde D_{\tilde m}\\[8pt]
  \tilde D_{\alpha}
 \end{pmatrix} = 
\begin{pmatrix*}[r]
 1 & 1 & 0 & -\frac{1}{2} \tensor{\cK}{_{\tilde 0}_{\tilde 0}^\beta} \\[8pt]
 0 & -1 & 0 & \tensor{\cK}{_{\tilde 0}_{\tilde 0}^\beta} \\[8pt]
 0 & -1 & \delta_{\tilde m}^{\tilde n} & \tensor{\cK}{_{\tilde 0}_{\tilde 0}^\beta} - \tensor{\cK}{_{\tilde 0}_{\tilde m}^\beta} \\[8pt]
 0 & 0 & 0 & \delta^\beta_\alpha
\end{pmatrix*}\cdot
\begin{pmatrix}
  D_{0}\\[8pt]
  D_{\tilde 0}\\[8pt]
  D_{\tilde n}\\[8pt]
  D_{\beta} \, 
 \end{pmatrix},
\end{align}
with the shorthand notation
\begin{subequations}
\begin{align}
 \tensor{\cK}{_\Lambda_\Sigma^\alpha} &:=\tensor{\eta}{^{-1}_\beta^\alpha}\ D_\Lambda \cdot D_\Sigma \cdot \cC^\beta   \qquad &&\textrm{three dimensions}, \\
 \tensor{\cK}{_\Lambda_\Sigma^\alpha} &:= D_\Lambda \cdot D_\Sigma \cdot D^\alpha   \qquad &&\textrm{five dimensions}.
\end{align}
\end{subequations}
It is easy to check that this is indeed the transformation induced by zero-section changes using the intersection numbers
listed in \autoref{sec:intersection_nos}.
Inserting the quantities from the classical matchings \eqref{e:cl_matching_3d}, \eqref{e:cl_matching_5d} for $\tensor{\cK}{_m_n^\alpha}$ into
\eqref{e:zero_section_change} and taking the transpose and inverse gives the transformation of the vectors in the field theory.
We point out that this map coincides with \eqref{e:abelian_trafo_vectors_4d}, \eqref{e:abelian_trafo_vectors_6d}
up to $U(1)$ basis transformations in the higher-dimensional theories.
Consequently the classical intersection numbers are only invariant up to $U(1)$ basis transformations performed already
in four and six dimensions.

We are now in the position to also identify the origin of the shifts \eqref{e:abelian_shift_4d} 
of the three-dimensional Chern-Simons terms. These were necessary in addition to the usual basis transformation 
when starting with four-dimensional pure Abelian models.
In the M-theory manifestation we can see that the expression for the Chern-Simons coefficients $\Theta_{\Lambda\Sigma}$
involves the $G_4$-flux \eqref{e:M-theory_CS}. The choice of flux is constrained by several conditions \cite{Grimm:2010ks,Marsano:2011hv,Grimm:2011sk,Cvetic:2012xn,Cvetic:2013uta}, one of them stating
$\Theta_{0\alpha}\overset{!}{=}0$, which guarantees the absence of circle fluxes in the dual setting. In order to preserve this condition under the map \eqref{e:zero_section_change} one also has to transform the flux according to
\begin{align}
 \tilde G_4 = G_4 - \tensor{\eta}{^{-1}_\beta^\alpha}\ (D_{\tilde 0} \cdot D_\alpha \cdot G_4 )\,\cC^\beta \, . 
\end{align}
We again stress that this transformation only appears for the Abelian models in four dimensions.
It is now easy to work out the anomaly conditions by using the transformations of the 
one-loop Chern-Simons terms as done in the previous sections.

The fact that zero-section changes reproduce all Abelian anomalies has far-reaching consequences: It states that
the invariance under the choice of the zero-section implies
the cancelation of Abelian anomalies in the effective theory of F-theory
compactifications on Calabi-Yau four- and threefolds. Hence this provides a proof for 
the absence of Abelian anomalies for F-theory compactifications on Calabi-Yau four- 
and threefolds considered here.

\subsection{Non-Abelian anomalies from zero-node changes} \label{zero-node-section}

In the following we describe the analog structure to the zero-section changes for non-Abelian gauge groups.
We consider the non-Abelian version of the divisor transformation \eqref{e:zero_section_change}
and try to find a geometric pattern that reproduces these maps. It turns out that this can be achieved by introducing a generalized base divisor
shift and a Shioda map for the Cartan divisors. These definitions depend on the choice of what we call the zero-node, and transformations of
divisors are induced by changes of zero-nodes. We restrict to a setting with a simple non-Abelian gauge group $G$ without Abelian factors and
exactly one section.

The non-Abelian version of the transformation \eqref{e:zero_section_change} is given by
\begin{align}\label{e:zero_node_change}
 \begin{pmatrix}
  \tilde D_{0}\\[8pt]
  \tilde D_{\tilde 0}\\[8pt]
  \tilde D_{\tilde I}\\[8pt]
  \tilde D_{\alpha}
 \end{pmatrix} = 
\begin{pmatrix*}[r]
 1 & 1 & 0 & -\frac{1}{2} \tensor{\cK}{_{\tilde 0}_{\tilde 0}^\beta} \\[8pt]
 0 & -1 & 0 & \tensor{\cK}{_{\tilde 0}_{\tilde 0}^\beta} \\[8pt]
 0 & -1 & \delta_{\tilde I}^{\tilde J} & \tensor{\cK}{_{\tilde 0}_{\tilde 0}^\beta} - \tensor{\cK}{_{\tilde 0}_{\tilde I}^\beta} \\[8pt]
 0 & 0 & 0 & \delta^\beta_\alpha
\end{pmatrix*}\cdot
\begin{pmatrix}
  D_{0}\\[8pt]
  D_{\tilde 0}\\[8pt]
  D_{\tilde J}\\[8pt]
  D_{\beta} \, 
 \end{pmatrix},
\end{align}
where we split $I=(\tilde 0 , \tilde I)$ for some arbitrary choice of $D_{\tilde 0}$.
It is possible to find a geometric interpretation of this transformation. Over the codimension-one locus in the base where the seven-brane sits
the fiber splits into $\rk G +1$ irreducible components. Fibering these over the seven-brane divisor in the base we obtain
$\rk G +1$ divisors of the total space which we denote by $\Sigma_0$, $\Sigma_I$. The choice of $\Sigma_0$ out of the $\rk G +1$ divisors
is arbitrary and we call it the zero-node. Having chosen a zero-node we can now write down a modified definition of the base divisor
\begin{subequations}\label{e:base_shifts_nA}
\begin{align}
\label{e:nA_base_shift_3d} D_0 = \sigma_0 & - \frac{1}{2} ( \sigma_0 \cdot \sigma_0 \cdot \cC^\alpha )\tensor{\eta}{^{-1}_\alpha^\beta} D_\beta 
 + (1 - \sigma_0 \cdot \Sigma_0 \cdot \cC ) 
 \Big [ \Sigma_0 - \frac{1}{2} ( \Sigma_0 \cdot \Sigma_0 \cdot \cC^\alpha )\tensor{\eta}{^{-1}_\alpha^\beta} D_\beta \Big ]  \nn \\
 & \qquad \textrm{three dimensions} \, ,\\
 \label{e:nA_base_shift_5d} D_0 = \sigma_0 &- \frac{1}{2} ( \sigma_0 \cdot \sigma_0 \cdot D^\alpha )D_\alpha 
 + (1 - \sigma_0 \cdot \Sigma_0 \cdot D ) 
 \Big [ \Sigma_0 - \frac{1}{2} ( \Sigma_0 \cdot \Sigma_0 \cdot D^\alpha )D_\alpha \Big ] \nn \\
  & \qquad \textrm{five dimensions} \, ,
\end{align}
\end{subequations}
as well as a Shioda map for the Cartan divisors\footnote{We assume in the following that the zero-section intersects an affine node of the Dynkin diagram,
which is certainly always true for toric constructions. We are confident that this generalizes to arbitrary geometries.}
\begin{subequations}\label{e:Shioda_shifts_nA}
\begin{align}
\label{e:nA_Shioda_3d} D_I = \Sigma_I  & + (1 - \sigma_0 \cdot \Sigma_0 \cdot \cC ) \Big [ - \Sigma_0 + \Big (\Sigma_0 \cdot (\Sigma_0 - \Sigma_I )
\cdot \cC^\alpha \Big )\tensor{\eta}{^{-1}_\alpha^\beta} D_\beta  \Big ] \nn \\
 &+ ( \sigma_0 \cdot \Sigma_I \cdot \cC )\Big [ - \Sigma_I
 + (\Sigma_0 \cdot \Sigma_I \cdot \cC^\alpha )\tensor{\eta}{^{-1}_\alpha^\beta} D_\beta \Big ] \qquad \textrm{three dimensions}, \\
\label{e:nA_Shioda_5d} D_I = \Sigma_I & + (1 - \sigma_0 \cdot \Sigma_0 \cdot D ) \Big [ - \Sigma_0 + \Big (\Sigma_0 \cdot (\Sigma_0 - \Sigma_I )
\cdot D^\alpha \Big ) D_\alpha  \Big ] \nn \\
 &+ ( \sigma_0 \cdot \Sigma_I \cdot D )\Big [ - \Sigma_I
 + (\Sigma_0 \cdot \Sigma_I \cdot D^\alpha ) D_\alpha \Big ] \qquad \textrm{five dimensions}\, .
\end{align}
\end{subequations}
For a Calabi-Yau fourfold $\hat Y$ we have introduced $\cC$ as the four-cycle $\pi^{-1}(\cC^b)$ in $\hat Y$ 
obtained from a curve $\cC^b$ in $B$ that intersects the seven-brane divisor exactly once.
Similarly, for a Calabi-Yau threefold $\hat Y$ we define $D$ to be the divisor $\pi^{-1}(D^b)$ in $\hat Y$, with $D^b$ being a divisor
in $B$ intersecting the seven-brane divisor precisely once.
The necessity to introduce $\cC$ and $D$ is known already from the Abelian Shioda map in the presence 
of non-Abelian singularities, their precise construction can be looked up in \cite{Morrison:2012ei}.
It is important that the expressions $(\sigma_0 \cdot \Sigma_0 \cdot \cC)$, $(\sigma_0 \cdot \Sigma_I \cdot \cC)$, $(\sigma_0 \cdot \Sigma_0 \cdot D)$,
$(\sigma_0 \cdot \Sigma_I \cdot D)$ equal to one if the respective node $\Sigma_0$, $\Sigma_I$ gets intersected by the zero-section $\sigma_0$, and
are zero otherwise.

We note that in the case that the zero-section intersects the zero-node we obtain the usual F-theory definitions
\begin{align}
 D_0 &= \sigma_0  - \frac{1}{2} ( \sigma_0 \cdot \sigma_0 \cdot \cC^\alpha )\tensor{\eta}{^{-1}_\alpha^\beta} D_\beta \quad &&\textrm{three dimensions}, \\
 D_0 &= \sigma_0  - \frac{1}{2} ( \sigma_0 \cdot \sigma_0 \cdot D^\alpha )D_\alpha \quad &&\textrm{five dimensions}, \\
 D_I &= \Sigma_I \, .
\end{align}
One can check that under changes of zero-nodes the geometric definitions \eqref{e:base_shifts_nA}, \eqref{e:Shioda_shifts_nA} induce
the transformations \eqref{e:zero_node_change}
on the divisors.
We stress that the corresponding transformation to \eqref{e:zero_node_change} of the vectors
in the circle reduced theory on the Coulomb branch is again the same as \eqref{e:non_abelian_trafo_vectors_4d}, \eqref{e:non_abelian_trafo_vectors_6d}
up to basis transformations solely among the Cartan generators, not involving the KK-vector.\footnote{Also
in \eqref{e:non_abelian_trafo_vectors_4d} we omitted the Green-Schwarz coefficients $b^\alpha$ because they drop out in the calculations
since $\Theta_{\alpha I}$ vanishes.} This means that anomalies are again canceled if \eqref{e:zero_node_change} is a symmetry of the theory.
Since independently of the chosen zero-node
the geometric definitions \eqref{e:base_shifts_nA}, \eqref{e:Shioda_shifts_nA} always reproduce the appropriate intersection numbers
that allow for a matching with the circle reduced theory on the
Coulomb branch, we claim that the theory is invariant under zero-node changes.
Invariance under zero-node changes then proves the absence of non-Abelian anomalies in F-theory compactifications
on Calabi-Yau four- and threefolds. 

We stress that for the completion of a rigorous proof of anomaly cancelation a deeper geometrical understanding of the definitions 
\eqref{e:base_shifts_nA} and \eqref{e:Shioda_shifts_nA} would be desirable. Our key observation is that making these definitions and 
performing the subsequent zero-node changes actually correspond to geometric symmetries of the Calabi-Yau geometry 
by inspecting the intersection numbers. This symmetry has not been exploited so far and appears much less 
intuitive than the corresponding zero-section exchanges in the Abelian case.

At this stage it might be fruitful to go once again through the notions \textit{zero-section, affine node, zero-node}
for clarification:
\begin{itemize}
 \item The choice of a particular \textit{zero-section} corresponds to the choice of the corresponding Weierstrass model,
 more precisely the zero-section is mapped to the $[ z=0 ]$-section of the Weierstrass model.
 \item The \textit{affine node} is precisely the node that is intersected by the zero-section.
 After the mapping to the (singular) Weierstrass model it is the only node which remains large, while all other nodes collapse to zero size.
 Importantly, by definition the affine node is linked to the choice of zero-section and cannot be picked independently.
 \item The \textit{zero-node} is some kind of auxiliary bookkeeping device of the different possibilities for taking the F-theory limit.
 It can be any node of the associated affine Dynkin diagram including the ones that are not intersected by any section. The geometric meaning
 of this freedom is not clear yet and will be subject of future investigations.
\end{itemize}

\begin{figure}
\centering
\begin{picture}(300,120)
 \includegraphics[scale=0.4]{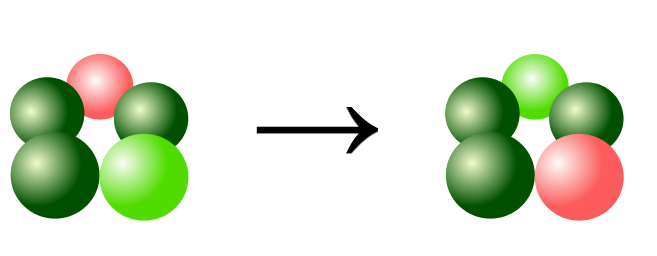}
 \put(-229,85){\large$\Sigma_0$}
 \put(-190,8){\large$\Sigma_{\tilde 0}$}
 \put(-247,35){\large$\sigma_0$}
 \put(-72,35){\large$\sigma_0$}
 \put(-255,28){$\bf\times$}
 \put(-80,28){$\bf\times$}
 \put(-16,8){\large$\tilde\Sigma_0$}
 \put(-53,85){\large$\tilde\Sigma_{\tilde 0}$}
\end{picture}
\caption{Two different choices for zero-nodes in a theory with $A_4$ gauge algebra are depicted. The zero-nodes are colored in red, the
remaining ones, corresponding to the Cartan generators, in green. Note that it is not necessary that the zero-node gets intersected by the zero-section.}
\label{p:zero_node_change}
\end{figure}
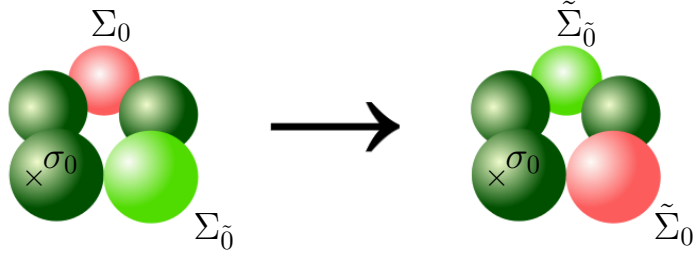

For illustration we depict a generic change of zero-nodes for an $A_4$-model in \autoref{p:zero_node_change}.
It is however crucial to realize that in models other than $A_n$ the zero-node does not have to be an affine node of the Dynkin diagram.
This is possible since the choice of a zero-node is just an auxiliary step in the non-Abelian Shioda map.
Indeed the affine node $\Sigma_{\textrm{aff}}$ in the end always drops out in the definitions of the Cartan divisors.
Let us illustrate this with an explicit calculation. Take a model with arbitrary gauge algebra and choose a zero-node $\Sigma_0$ which cannot be interpreted
as an affine node of the Dynkin diagram. We then first apply the Shioda map to all other nodes that are not intersected by the zero-section. We find
e.g.~in six dimensions
\begin{align}
 D_I = \Sigma_I - \Sigma_0 + \Big ( \Sigma_0 \cdot (\Sigma_0 - \Sigma_I ) \cdot D^\alpha \Big ) D_\alpha \, .
\end{align}
Finally there is still the node left which is intersected by the zero-section, denoted by $\Sigma_{\textrm{aff}}$.
Although, as already stated before this is an affine node of the Dynkin diagram,
it however also defines a Cartan divisor via the Shioda map
\begin{align}
 D_{\textrm{aff}} = - \Sigma_0 + (\Sigma_0 \cdot \Sigma_0 \cdot D^\alpha ) D_\alpha \, .
\end{align}
Note that the notation $D_{\textrm{aff}}$ means just that this divisor is derived from the affine node, as a Cartan divisor it still belongs
to a simple Lie-algebra.
Importantly one realizes that the affine node $\Sigma_{\textrm{aff}}$ finally drops out of all definitions, which is somehow expected.

\section{Conclusions}
In this paper we studied symmetries of four- and six-dimensional matter coupled gauge theories
compactified on a circle. The considered transformations act on the circle-compactified theories
by mixing the Kaluza-Klein vector arising when reducing the metric with the Kaluza-Klein zero-mode of a Cartan gauge field of 
a non-Abelian group or a $U(1)$ Abelian gauge field. They are induced by a large gauge 
transformation in the higher-dimensional theory with support in the circle direction. 
Under this action the whole Kaluza-Klein tower of fields gets rearranged.  
If the higher-dimensional theory does not cancel anomalies, this 
has profound implications for the lower-dimensional one-loop 
effective theory for the massless modes. The considered 
transformations allow us to exactly 
extract the anomaly conditions from the effective theory.  

Our focus was on the Chern-Simons terms in the 
three- and five-dimensional effective theory that are known to receive one-loop 
corrections when integrating out massive fields, which are precisely 
the Kaluza-Klein modes of higher-dimensional fields that contribute to the anomaly. 
Their mass depends on the Coulomb branch parameters
and the circle radius if they are excited Kaluza-Klein modes. Depending 
on the Coulomb branch parameters and the circle radius there seemingly 
exists a discrete set of infinitely many effective theories for the massless modes 
differing by their one-loop Chern-Simons terms. 
We have shown that precisely when anomalies are canceled 
the transformations mixing gauge fields and the Kaluza-Klein vector identify 
identical effective theories. This is in accord with the fact that large gauge 
transformations are only symmetries of the one-loop quantum theory 
if anomalies are canceled. In other words, the naive moduli space of the Coulomb 
branch parameters in an anomaly free theory can be modded out by the shifts 
\eqref{shift-vevs} to label inequivalent theories.  
Note that although it was already known that Chern-Simons terms know about higher-dimensional 
anomalies, the direct relation to the anomaly conditions and the periodicity of the 
Coulomb branch parameters seems to have not appeared in 
the literature before. 

A closer look on the considered symmetry transformations shows 
that they act in a rather non-trivial way in the Kaluza-Klein perspective. 
For example, for a non-Abelian gauge group already within a single representation of 
matter fields the shifts of the Kaluza-Klein level act differently on the fields depending 
on the weight of the field in the representation. Furthermore, this is accompanied with the 
fact that in non-Abelian gauge theories the gauge transformation of the transformed 
Kaluza-Klein vector is not anymore that of a $U(1)$ field but it rather mixes with the 
non-Abelian gauge transformations. All these modification have to be taken into account 
when computing the one-loop effective theory. In addition, we have found 
that in four-dimensional theories with $U(1)$ gauge factors and Green-Schwarz 
axions an additional shift of the Chern-Simons terms has to be performed.
This shift ensures the absence of circle fluxes for the axions that otherwise 
would complicate the one-loop computation. Nevertheless, it would be interesting 
to evaluate the effective theory for this more general situation directly.    

The original motivation for the study of the symmetries 
came from the M-/F-theory duality. More precisely F-theory on an elliptically fibered Calabi-Yau
manifold compactified on an additional circle and pushed to the Coulomb branch is dual to M-theory on the resolved Calabi-Yau space.
The effective action of the F-theory compactification is then obtained from matching a circle reduced supergravity theory with the M-theory
compactification. This implies that a study of anomaly cancelation for F-theory effective actions requires to approach them 
using a lower-dimensional perspective. Up to this work, anomaly cancelation was mostly studied on a case-by-case basis, by extracting 
the spectrum and Green-Schwarz terms for a given geometry and background flux. All studied examples were shown to be anomaly 
free. Only when restricting to special geometries and a subset of anomalies it was possible to show anomaly cancelation generally. 
In this work we proposed a general argument that geometric symmetries actually ensure cancelation of anomalies for 
a general class of Calabi-Yau manifolds. 

In F-theory compactifications with only Abelian gauge symmetries there is a clear identification of 
the transformation mixing the Kaluza-Klein vector and the $U(1)$ gauge fields. Recall that the 
Kaluza-Klein vector corresponds to the zero-section of the elliptic fibration while further $U(1)$ gauge fields
correspond to additional sections of the geometry. However, since the M-theory to F-theory duality 
does not depend on the choice of the zero-section one is free to pick any section of the fibration. 
It turned out that the different choices of zero-sections exactly induce the 
transformations encountered in the study of anomalies. Since this is actually a geometrical 
symmetry this shows the general cancelation of pure Abelian anomalies in 
F-theory compactifications on Calabi-Yau manifolds. Note that we found that 
changing the choice of the zero-section implies that the four-form flux $G_4$ has 
to be shifted in order to avoid the appearance of circle fluxes. Generalizing the 
the one-loop computation to include circle fluxes we believe that the same results 
can be obtained without shifting $G_4$. This implies that more general 
classes of $G_4$ might be allowed in F-theory. 

For non-Abelian gauge theories in F-theory the identification of an analog 
geometric symmetry turned out to be more involved. Nevertheless, we were able 
to propose a geometric symmetry, corresponding to zero-node changes, following 
our field theory insights. Over a seven-brane divisor in the fully resolved geometry 
the fiber splits into irreducible components intersecting
as the affine Dynkin diagram of the gauge algebra. One of these 
components is chosen to be the zero-node.
Using all components and the picked zero-node 
we introduced a refined definition of divisors corresponding to the Cartan elements 
of the gauge group and the Kaluza-Klein vector. These were shown to leave the  
classical intersection numbers invariant up to a higher-dimensional basis transformation.
We identified our symmetry transformation as the freedom to pick a zero-node. 
This shows the general absence of non-Abelian gauge anomalies
in F-theory compactifications on Calabi-Yau manifolds.
It would be interesting to gain a better geometric understanding for the freedom of picking a zero-node
and the related refined constructions of divisors.  Furthermore, it would also be desirable to generalize the 
analysis of phases of lower-dimensional gauge theories of 
\cite{Grimm:2011fx,Hayashi:2013lra,Hayashi:2014kca,Esole:2014bka,Esole:2014hya,Braun:2014kla} 
to fully include the Kaluza-Klein vector.

Let us comment on possible generalizations of our findings. It is an exciting 
question whether our approach generalizes to more complicated compactifications of consistent 
supergravity theories or string theory. 
Moreover, it would be interesting to study other anomalies in a similar spirit. 
We are confident that our perspective 
straightforwardly generalizes to the treatment of mixed Abelian-non-Abelian
anomalies. More involved are purely gravitational anomalies, in 
which case one would expect that the mixing of the Kaluza-Klein 
vector with the spin connection needs to be studied. Finally, it would be very interesting  
to address anomalies for other symmetries, such as conformal anomalies or 
R-symmetry anomalies.

\subsubsection*{Acknowledgments}
We would like to thank Federico Bonetti, Jan Keitel, and Eugenio Meg\'ias for illuminating discussions
and very useful remarks on the draft. This work was supported by a grant of the Max Planck Society.

\appendix

\section{Identities on the Coulomb branch}\label{CB_id}
In this section we show the central identities
\begin{align}
 q_{\tilde 0} &= \big(l_{q} + \frac{1}{2}\big) \sign (m^{q}_{\rm CB}) - \big(\tilde l_q + \frac{1}{2}\big) \sign (\tilde m^q_{\rm CB}) \, ,  \\
 w_{\tilde 0} &= \big(l_{w}+ \frac{1}{2}\big) \sign (m^{w}_{\rm CB}) - \big(\tilde  l_{w} + \frac{1}{2}\big) \sign (\tilde m^w_{\rm CB}) \, .
\end{align}
under the transformations \eqref{e:non_abelian_trafo_vectors_4d}, \eqref{e:abelian_trafo_vectors_4d},
\eqref{e:non_abelian_trafo_vectors_6d}, \eqref{e:abelian_trafo_vectors_6d}.
We only prove the first identity, since the second one works out in exactly the same way.
Consider a massive mode with charge vector $q$ under Abelian gauge bosons $A^m$, Coulomb branch mass $m^q_{\rm CB}$ and KK-level $n$. 
We pick one $A^{\tilde 0}$ and perform one of the basis changes \eqref{e:abelian_trafo_vectors_4d}, \eqref{e:abelian_trafo_vectors_6d}.
It is important to notice that these transformations leave all VEVs invariant except of
\begin{align}
 \langle \zeta^{\tilde 0} \rangle \mapsto \langle \zeta^{\tilde 0} \rangle - \frac{1}{\langle r \rangle} \, .
\end{align}
One can then easily show that the sign function fulfills
\begin{align}\label{e:sign_trafo}
\sign (m^q_{\rm CB} + n \, m_{\rm KK}) = \sign (\tilde  m^{q}_{\rm CB} + ( n + q_{\tilde 0})\, m_{\rm KK}) \, .
\end{align}
Depending on the sign of the Coulomb branch masses we have to investigate four different cases:
\begin{enumerate}
 \item \underline{$\sign (m^q_{\rm CB}) > 0 $}
 
 The integer quantity $l_q$ is then defined via the following property
 \begin{align}
  \sign (m^q_{\rm CB} - l_q \, m_{\rm KK}) > 0 \qquad \wedge \qquad \sign (m^q_{\rm CB} - ( l_q + 1 ) \, m_{\rm KK}) < 0 \, .
 \end{align}
 Using \eqref{e:sign_trafo} we find
 \begin{align}
  \sign (\tilde m^{q}_{\rm CB} - (l_q - q_{\tilde 0}) \, m_{\rm KK}) > 0 \qquad \wedge \qquad \sign (\tilde m^{q}_{\rm CB} -
  (l_q - q_{\tilde 0}+1) \, m_{\rm KK}) < 0 \, .
 \end{align}
 Depending on the sign of $m^{\tilde q}_{\rm CB}$ we can now read off $\tilde l_{q}$
 \begin{subequations}\label{e:cases1} 
 \begin{align}
 &\tilde  l_{q} = l_q - q_{\tilde 0}  && \textrm{for } \sign(\tilde m^{q}_{\rm CB}) > 0 \\
& \tilde l_{q} = -l_q + q_{\tilde 0}-1  && \textrm{for } \sign(\tilde m^{q}_{\rm CB}) < 0 \, .
 \end{align}
 \end{subequations}
\item \underline{$\sign (m^q_{\rm CB}) < 0 $}

 Now $l_q$ is defined as
 \begin{align}
  \sign (m^q_{\rm CB} + ( l_q + 1 ) \, m_{\rm KK}) > 0 \qquad \wedge \qquad \sign (m^q_{\rm CB} + l_q   \, m_{\rm KK}) < 0 \, .
 \end{align}
 With \eqref{e:sign_trafo} we get
 \begin{align}
  \sign (\tilde m^{q}_{\rm CB} + (l_q + q_{\tilde 0} +1 ) \, m_{\rm KK}) > 0
  \qquad \wedge \qquad \sign (\tilde m^{q}_{\rm CB} + (l_q + q_{\tilde 0}) \, m_{\rm KK}) < 0 \, .
 \end{align}
 From this we can again determine $\tilde l_{q}$
 \begin{subequations}\label{e:cases2}
 \begin{align}
& \tilde l_{q} = - l_q - q_{\tilde 0} -1  && \textrm{for } \sign(\tilde m^{q}_{\rm CB}) > 0 \\
&\tilde l_{q} = l_q + q_{\tilde 0}  && \textrm{for } \sign(\tilde m^{q}_{\rm CB}) < 0 \, .
 \end{align}
 \end{subequations}
\end{enumerate}
It is now easy to check that the relations \eqref{e:cases1}, \eqref{e:cases2} are summarized as
\begin{align}
 q_{\tilde 0} &= \big(l_{q} + \frac{1}{2}\big) \sign (m^{q}_{\rm CB}) - \big(\tilde l_q + \frac{1}{2}\big) \sign (\tilde m^q_{\rm CB}) \, .
\end{align}
In complete analogy one proves the identity
\begin{align}
 w_{\tilde 0} &= \big(l_{w}+ \frac{1}{2}\big) \sign (m^{w}_{\rm CB}) - \big( \tilde l_{w} + \frac{1}{2}\big) \sign (\tilde m^w_{\rm CB}) \, .
\end{align}

Finally considering the basis transformations in the M-/F-theory setting \eqref{e:zero_section_change}, \eqref{e:zero_node_change}
one can show in the same way
\begin{align}
 q_{\tilde 0} &= - \big(l_{q} + \frac{1}{2}\big) \sign (m^{q}_{\rm CB}) + \big(\tilde l_q + \frac{1}{2}\big) \sign (\tilde m^q_{\rm CB}) \, , \\
 w_{\tilde 0} &= - \big(l_{w}+ \frac{1}{2}\big) \sign (m^{w}_{\rm CB}) + \big( \tilde l_{w} + \frac{1}{2}\big) \sign (\tilde m^w_{\rm CB}) \, .
\end{align}
Note the sign change compared to the relations before.

\section{One-loop calculations}\label{app:loops}
In this section we perform the loop-calculations to find the corrections to the Chern-Simons terms.
Around \eqref{e:4d_single_loop_CS} it is noted that in three dimensions one-loop Chern-Simons terms arise from spin-$\sfrac{1}{2}$ fermions,
the contribution of a single Dirac field given by \cite{Niemi:1983rq,Redlich:1983dv,Aharony:1997bx}
\begin{align}
 \Theta_{\Lambda\Sigma}^{\textrm{loop}} = \frac{1}{2} q_\Lambda q_\Sigma \sign (m) \, .
\end{align}
In five dimensions one-loop Chern-Simons terms are induced by spin-$\sfrac{1}{2}$ and spin-$\sfrac{3}{2}$ fermions as well as two-forms, as already stated around
\eqref{e:6d_single_field_CS_1}, \eqref{e:6d_single_field_CS_2}. The corrections originating from a single
field are \cite{Witten:1996qb,Bonetti:2012fn,Bonetti:2013ela}
\begin{align}
k_{\Lambda\Sigma\Theta}^{\textrm{loop}} = c_{AFF}\, q_\Lambda q_\Sigma q_\Theta \, \sign (m) \, , \\
k_{\Lambda}^{\textrm{loop}} = c_{A\cR\cR}\, q_\Lambda \, \sign (m) \, .
\end{align}
The quantities $c_{AFF}$, $c_{A\cR\cR}$ are listed in \autoref{t:CS_correct}.
Since we are treating circle compactified theories in this paper, the full contributions to the one-loop Chern-Simons terms are generically
infinite sums over KK-modes, which need to be treated with zeta function regularization. In these calculations four different
types of sums do appear in general:
\begin{align}
\sum_{n = -\infty}^{+\infty}\sign (x+n)\, , \qquad &\sum_{n = -\infty}^{+\infty}n\sign (x+n)\, ,\\
\sum_{n = -\infty}^{+\infty}n^2 \sign (x+n)\, , \qquad &\sum_{n = -\infty}^{+\infty}n^3 \sign (x+n) \, . \nn
\end{align}
Using zeta function regularization
\begin{align}
 \sum_{n = 1}^{\infty}n &\mapsto \zeta(-1) = - \frac{1}{12} \, , 
 &\sum_{n = 1}^{\infty}n^3 &\mapsto \zeta(-3) = \frac{1}{120} \, .
\end{align}
these sums become
{\allowdisplaybreaks\begin{align}
 \sum_{n = -\infty}^{+\infty}\sign (x+n) &= (2l + 1)\sign (x) \, , \\
 \sum_{n = -\infty}^{+\infty} n \sign (x+n) &= - \frac{1}{6} - l(l+1) \, , \\
 \sum_{n = -\infty}^{+\infty} n^2 \sign (x+n) &= \frac{1}{3}\, l(l+1)(2l + 1)\sign (x) \, , \\
 \sum_{n = -\infty}^{+\infty} n^3 \sign (x+n) &= \frac{1}{60} - \frac{1}{2}\, l^2 (l+1)^2 \, ,
\end{align}}
where
\begin{align}
 l := \big \lfloor \vert x \vert \big \rfloor \, ,
\end{align}
making use of the floor function $\lfloor \cdot \rfloor$.

In order to find the full set of one-loop Chern-Simons coefficients for four-dimensional theories on the circle as introduced in \autoref{sec:4d_anomalies}
we note that four-dimensional Weyl fermions reduce to three-dimensional Dirac fermions with
$\sign (m)= \sign(m_{\textrm{CB}}+ n m_{\textrm{KK}} )$ or $\sign (m)= -\sign(m_{\textrm{CB}}+ n m_{\textrm{KK}}$
for a former left-handed or right-handed spinor, respectively.
We evaluate for the pure non-Abelian theory \cite{Grimm:2013oga,Cvetic:2013uta}
\begin{subequations}
\begin{align}
 \Theta_{00}& = \frac{1}{3} \sum_{R} F(R) \sum_{w \in R} \ l_w \ (l_w +1 ) \ (l_w +\frac{1}{2} ) \ \sign  (m^w_{\rm CB})\, ,\\
\Theta_{0I} &= \frac{1}{12}\sum_{R} F(R) \sum_{w \in R} \ \big(1+ 6\ l_w \ (l_w +1 )\big) \ w_I\, , \\
 \Theta_{IJ} &=  \sum_{R} F(R) \sum_{w \in R}  \ (l_w +\frac{1}{2} ) \ w_I w_J \ \sign  (m^w_{\rm CB}) \, ,
\end{align}
\end{subequations}
where the sums are over all representations and all weights of a given representation.
In the pure Abelian theory we obtain
\begin{subequations}
\begin{align}
\Theta_{00}& = \frac{1}{3} \sum_q F(q)  \ l_q \ (l_q +1 ) \ (l_q +\frac{1}{2} ) \ \sign  (m^q_{\rm CB})\, ,\\
 \Theta_{0m} &= \frac{1}{12}\sum_{q}  F(q) \ \big(1+ 6\ l_q \ (l_q +1 )\big) \ q_m\, , \\
 \Theta_{mn} &=  \sum_{q}  F(q) \ (l_q +\frac{1}{2} ) \ q_m q_n \ \sign  (m^q_{\rm CB}) \, .
\end{align}
\end{subequations}

In six-dimensional theories on the circle following the pattern of \autoref{sec:6d_anomalies}
we realize that Weyl spinors reduce to Dirac spinors and the KK-modes of former (anti-)self-dual tensors
are massive two-forms with first order kinetic terms, the corresponding Lagrangians can be looked up e.g.~in \cite{Bonetti:2013ela,Grimm:2014soa,Grimm:2014aha}.
The contributions of these fields to the loop-corrections can then be inferred from \autoref{t:CS_correct}.
We note that $\sign (m)$ on these modes reads
\begin{align}
 \sign (m) = \begin{cases}
              + \sign(m_{\textrm{CB}}+ n \, m_{\textrm{KK}} ) \quad \textrm{for } (\frac{1}{2},0),(1,0),(1,\frac{1}{2})\, , \\
              - \sign(m_{\textrm{CB}}+ n \, m_{\textrm{KK}} )\quad \textrm{for } (0,\frac{1}{2}),(0,1),(\frac{1}{2},1) \, ,
             \end{cases}
\end{align}
where we labeled the representations of the former six-dimensional massless fields under the massless little group in six dimensions $SU(2) \times SU(2)$
by their spins. 
Furthermore note that because of the (anti-)self-duality condition of the tensors in six dimensions the contribution of a corresponding KK-mode
is only half the one listed in \autoref{t:CS_correct}.\footnote{This is also the case if one reduces Majorana-Weyl
rather than Weyl spinors on a circle to five dimensions.}
Finally the corrections for the pure non-Abelian theory
are given by \cite{Grimm:2013oga}
{\allowdisplaybreaks\begin{subequations}
\begin{align}
 k_{000} &= \frac{1}{120} \Big (2(T_{sd} - T_{asd})-F_{\fe}-5 F_{\gr}\Big ) 
 + \frac{1}{4}\sum_R F_{\fe}(R) \sum_{w \in R} \ l_w^2 \ (l_w +1 )^2  \, ,\\
k_{00I} &=  \frac{1}{6}\sum_R F_{\fe}(R) \sum_{w \in R} \ l_w \ (l_w +1 ) \ (2l_w +1 ) \ w_I \ \sign  (m^w_{\rm CB})\, ,\\
k_{0IJ} &=  \frac{1}{12}\sum_R F_{\fe}(R) \sum_{w \in R} \ (1+ 6\ l_w \ (l_w +1 )) \ w_I w_J\, ,\\
 k_{IJK} &= \frac{1}{2}\sum_R F_{\fe}(R) \sum_{w \in R} \ (2l_w +1 ) \ w_I w_J w_K \ \sign  (m^w_{\rm CB})\, ,\\
k_{0} &= \frac{1}{6} \Big (19 F_{\gr} - F_{\fe} -4(T_{sd} - T_{asd})\Big ) - \sum_R F_{\fe}(R) \sum_{w \in R} \ l_w \ (l_w +1 )\, , \\
 k_{I} &= -\sum_R F_{\fe}(R) \sum_{w \in R} \ (2l_w +1 ) \ w_I \ \sign  (m^w_{\rm CB}) \, ,
\end{align}
\end{subequations}}
and for the Abelian theory they read
{\allowdisplaybreaks\begin{subequations}
\begin{align}
k_{000} &= \frac{1}{120} \Big (2(T_{sd} - T_{asd})-F_{\fe}-5 F_{\gr}\Big ) + \frac{1}{4}\sum_{q} F_{\fe}(q) \ l_q^2 \ (l_q +1 )^2 \, , \\
k_{00m} &=  \frac{1}{6}\sum_{q}  F_{\fe}(q) \ l_q \ (l_q +1 ) \ (2l_q +1 ) \ q_m \ \sign  (m^q_{\rm CB})\, ,\\
k_{0mn} &=  \frac{1}{12}\sum_{q}  F_{\fe}(q) \ \big(1+ 6\ l_q \ (l_q +1 )\big) \ q_m q_n\, ,\\
k_{mnp} &= \frac{1}{2}\sum_{q}  F_{\fe}(q) \ (2l_q +1 ) \ q_m q_n q_p \ \sign  (m^q_{\rm CB})\, ,\\
k_{0} &= \frac{1}{6} \Big (19 F_{\gr} - F_{\fe} -4(T_{sd} - T_{asd})\Big ) - \sum_{q} F_{\fe}(q) \ l_q \ (l_q +1 ) \, , \\
k_{m} &= -\sum_{q}  F_{\fe}(q) \ (2l_q +1 ) \ q_m \ \sign  (m^q_{\rm CB}) \, .
\end{align}
\end{subequations}}

\section{Lie theory conventions and trace identities}\label{app:traces}
In this section we summarize our conventions for the Lie algebra theory in this paper.
Furthermore we show that the factors appearing in trace reductions of some representation of a simple Lie algebra can be related to
the sum over the weights in that representation. This will allow us to relate one-loop Chern-Simons terms to non-Abelian anomaly
cancelation conditions
since in the former sums over weights are evaluated, while in the latter factors of trace reductions appear.

Consider a simple Lie algebra $\mathfrak{g}$. We define a (preliminary) basis of Cartan generators $\lbrace\tilde T_i \rbrace$ enforcing
\begin{align}
 \tr_f ( \tilde T_i \tilde T_j ) = \delta_{ij} \, ,
\end{align}
which encodes the normalization of the root lattice, and the trace $\tr_f$ is taken in the fundamental representation.
We denote the simple roots by $\boldsymbol{\alpha}_I$, $I=1,\dots ,\rk \mathfrak{g}$, the simple coroots
are denoted by $\boldsymbol{\alpha}_I^\vee := \frac{2 \boldsymbol{\alpha}_I}{\langle \boldsymbol{\alpha}_I , \boldsymbol{\alpha}_I \rangle}$.
It turns out that for the considerations in this paper the coroot-basis $\lbrace T_I \rbrace$ for the Cartan-subalgebra is more convenient, it is given by
\begin{align}
 T_I := \frac{2\,\boldsymbol{\alpha}_I^i \tilde T_i}{\langle\boldsymbol{\alpha}_I ,\boldsymbol{\alpha}_I \rangle}
\end{align}
with $\boldsymbol{\alpha}_I^i$ the components of the simple roots.
We furthermore define the (normalized)
coroot intersection matrix $\cC_{IJ}$ as
\begin{align}\label{e:def_coroot_int_mat}
 \cC_{IJ} = \frac{1}{\lambda (\mathfrak{g})} \langle \boldsymbol{\alpha}^\vee_I , \boldsymbol{\alpha}^\vee_J \rangle \, ,
\end{align}
with
\begin{align}
 \lambda (\mathfrak{g}) = \frac{2}{\langle \boldsymbol{\alpha}_{\textrm{max}}, \boldsymbol{\alpha}_{\textrm{max}} \rangle} \, ,
\end{align}
where $\boldsymbol{\alpha}_{\textrm{max}}$ is the root of maximal length.
The normalization of the Cartan generators $T_I$ (in the coroot basis) can then be checked to take the form
\begin{align}
 \tr_f (T_I T_J) = \lambda (\mathfrak{g}) \, \cC_{IJ} \, .
\end{align}
Furthermore for some weight $w$ the Dynkin labels are defined as
\begin{align}
 w_I := \langle \boldsymbol{\alpha}^\vee_I , w \rangle \, .
\end{align}
Finally in \autoref{t:lie_conventions} we display the numbering of the nodes in the Dynkin diagrams, the definition of the fundamental
representations of all simple Lie algebras, as well as the values for the normalization factors $\lambda (\mathfrak{g})$ in our conventions.
\begin{table}
\begin{center}
\begin{tabular}{|m{1.4cm}||m{5.3cm}|m{5cm}|m{0.8cm}|}
\hline
algebra & Dynkin diagram & fundamental representation & $\lambda (\mathfrak{g})$\\
\hline \hline
$A_n$ & \scalebox{3}{\begin{dynkin}\tikzstyle{every node}=[font=\tiny\tiny]
    \node[align=left, scale=0.5] at (\dynkinstep*1,-.15cm){1};
    \node[align=left, scale=0.5] at (\dynkinstep*2,-.15cm){2};
    \node[align=left, scale=0.5] at (\dynkinstep*3,-.15cm){3};
    \node[align=left, scale=0.5] at (\dynkinstep*6,-.15cm){n-1};
    \node[align=left, scale=0.5] at (\dynkinstep*7,-.162cm){n};
    \node[align=left, scale=0.5] at (\dynkinstep*1,.1cm){};
    \dynkinline{1}{0}{4}{0};
    \dynkindots{4}{0}{5}{0};
    \dynkinline{5}{0}{7}{0};
    \foreach \x in {1,2,3,6,7}
    {
       \dynkindot{\x}{0}
    }
  \end{dynkin}} & $(1,0,0,\dots,0,0)$ & 1\\
\hline 
$B_n$ & \scalebox{3}{\begin{dynkin}\tikzstyle{every node}=[font=\tiny]
    \node[align=left, scale=0.5] at (\dynkinstep*1,-.15cm){1};
    \node[align=left, scale=0.5] at (\dynkinstep*2,-.15cm){2};
    \node[align=left, scale=0.5] at (\dynkinstep*3,-.15cm){3};
    \node[align=left, scale=0.5] at (\dynkinstep*6,-.15cm){n-1};
    \node[align=left, scale=0.5] at (\dynkinstep*7,-.162cm){n};
    \node[align=left, scale=0.5] at (\dynkinstep*1,.1cm){};
    \dynkinline{1}{0}{4}{0};
    \dynkindots{4}{0}{5}{0};
    \dynkinline{5}{0}{6}{0}
    \dynkindoubleline{6}{0}{7}{0};
    \foreach \x in {1,2,3,6,7}
    {
       \dynkindot{\x}{0}
    }
  \end{dynkin}} & $(1,0,0,\dots,0,0)$ & 2\\ 
\hline 
$C_n$ & \scalebox{3}{\begin{dynkin}\tikzstyle{every node}=[font=\tiny]
    \node[align=left, scale=0.5] at (\dynkinstep*1,-.15cm){1};
    \node[align=left, scale=0.5] at (\dynkinstep*2,-.15cm){2};
    \node[align=left, scale=0.5] at (\dynkinstep*3,-.15cm){3};
    \node[align=left, scale=0.5] at (\dynkinstep*6,-.15cm){n-1};
    \node[align=left, scale=0.5] at (\dynkinstep*7,-.162cm){n};
    \node[align=left, scale=0.5] at (\dynkinstep*1,.1cm){};
    \dynkinline{1}{0}{4}{0};
    \dynkindots{4}{0}{5}{0};
    \dynkinline{5}{0}{6}{0}
    \dynkindoubleline{7}{0}{6}{0};
    \foreach \x in {1,2,3,6,7}
    {
       \dynkindot{\x}{0}
    }
  \end{dynkin}} & $(1,0,0,\dots,0,0)$ & 1\\
\hline 
$D_n$ & \scalebox{3}{\begin{dynkin}\tikzstyle{every node}=[font=\tiny]
    \node[align=left, scale=0.5] at (\dynkinstep*1,-.15cm){1};
    \node[align=left, scale=0.5] at (\dynkinstep*2,-.15cm){2};
    \node[align=left, scale=0.5] at (\dynkinstep*3,-.15cm){3};
    \node[align=left, scale=0.5] at (\dynkinstep*6,-.15cm){n-2};
    \node[align=left, scale=0.5] at (\dynkinstep*7,-.35cm){n-1};
    \node[align=left, scale=0.5] at (\dynkinstep*7,.35cm){n};
    \dynkinline{1}{0}{4}{0};
    \dynkindots{4}{0}{5}{0};
    \dynkinline{5}{0}{6}{0}
    \foreach \x in {1,2,3,6}
    {
       \dynkindot{\x}{0}
    }
    \dynkindot{7}{.8}
    \dynkindot{7}{-.8}
    \dynkinline{6}{0}{7}{.8}
    \dynkinline{6}{0}{7}{-.8}
  \end{dynkin}} & $(1,0,0,\dots,0,0,0)$ & 2\\
\hline 
$E_6$ & \scalebox{3}{\begin{dynkin}\tikzstyle{every node}=[font=\tiny]
    \node[align=left, scale=0.5] at (\dynkinstep*1,-.15cm){1};
    \node[align=left, scale=0.5] at (\dynkinstep*2,-.15cm){3};
    \node[align=left, scale=0.5] at (\dynkinstep*3,-.15cm){4};
    \node[align=left, scale=0.5] at (\dynkinstep*4,-.15cm){5};
    \node[align=left, scale=0.5] at (\dynkinstep*5,-.15cm){6};
    \node[align=left, scale=0.5] at (\dynkinstep*3.5,.25cm){2};
    \foreach \x in {1,...,5}
    {
        \dynkindot{\x}{0}
    }
    \dynkindot{3}{1}
    \dynkinline{1}{0}{5}{0}
    \dynkinline{3}{0}{3}{1}
  \end{dynkin}} & $(0,0,0,0,0,1)$ & 6\\
\hline 
$E_7$ & \scalebox{3}{\begin{dynkin}\tikzstyle{every node}=[font=\tiny]
    \node[align=left, scale=0.5] at (\dynkinstep*1,-.15cm){1};
    \node[align=left, scale=0.5] at (\dynkinstep*2,-.15cm){3};
    \node[align=left, scale=0.5] at (\dynkinstep*3,-.15cm){4};
    \node[align=left, scale=0.5] at (\dynkinstep*4,-.15cm){5};
    \node[align=left, scale=0.5] at (\dynkinstep*5,-.15cm){6};
    \node[align=left, scale=0.5] at (\dynkinstep*6,-.15cm){7};
    \node[align=left, scale=0.5] at (\dynkinstep*3.5,.25cm){2};
    \foreach \x in {1,...,6}
    {
        \dynkindot{\x}{0}
    }
    \dynkindot{3}{1}
    \dynkinline{1}{0}{6}{0}
    \dynkinline{3}{0}{3}{1}
  \end{dynkin}} & $(0,0,0,0,0,0,1)$ & 12\\
\hline 
$E_8$ & \scalebox{3}{\begin{dynkin}\tikzstyle{every node}=[font=\tiny]
    \node[align=left, scale=0.5] at (\dynkinstep*1,-.15cm){1};
    \node[align=left, scale=0.5] at (\dynkinstep*2,-.15cm){3};
    \node[align=left, scale=0.5] at (\dynkinstep*3,-.15cm){4};
    \node[align=left, scale=0.5] at (\dynkinstep*4,-.15cm){5};
    \node[align=left, scale=0.5] at (\dynkinstep*5,-.15cm){6};
    \node[align=left, scale=0.5] at (\dynkinstep*6,-.15cm){7};
    \node[align=left, scale=0.5] at (\dynkinstep*7,-.15cm){8};
    \node[align=left, scale=0.5] at (\dynkinstep*3.5,.25cm){2};
    \foreach \x in {1,...,7}
    {
        \dynkindot{\x}{0}
    }
    \dynkindot{3}{1}
    \dynkinline{1}{0}{7}{0}
    \dynkinline{3}{0}{3}{1}
  \end{dynkin}} & $(0,0,0,0,0,0,0,1)$ & 60\\
\hline 
$F_4$ &  \scalebox{3}{\begin{dynkin}\tikzstyle{every node}=[font=\tiny]
    \node[align=left, scale=0.5] at (\dynkinstep*1,-.15cm){1};
    \node[align=left, scale=0.5] at (\dynkinstep*2,-.15cm){2};
    \node[align=left, scale=0.5] at (\dynkinstep*3,-.15cm){3};
    \node[align=left, scale=0.5] at (\dynkinstep*4,-.15cm){4};
    \node[align=left, scale=0.5] at (\dynkinstep*1,.1cm){};
    \dynkindoubleline{2}{0}{3}{0}
    \foreach \x in {1,...,4}
    {
        \dynkindot{\x}{0}
    }
    \dynkinline{1}{0}{2}{0}
    \dynkinline{3}{0}{4}{0}
  \end{dynkin}}  & $(0,0,0,1)$ & 6\\
\hline 
$G_2$ &  \scalebox{3}{\begin{dynkin}\tikzstyle{every node}=[font=\tiny]
    \node[align=left, scale=0.5] at (\dynkinstep*1,-.15cm){1};
    \node[align=left, scale=0.5] at (\dynkinstep*2,-.15cm){2};
    \node[align=left, scale=0.5] at (\dynkinstep*1,.1cm){};
    \dynkintripleline{2}{0}{1}{0}
    \foreach \x in {1,2}
    {
        \dynkindot{\x}{0}
    }
  \end{dynkin}}  & $(1,0)$ & 2 \\
\hline
\end{tabular}
\end{center}
\caption{Conventions for the simple Lie algebras.}
\label{t:lie_conventions}
\end{table}

\subsection{Cubic trace identities}
We now show that the conditions
\begin{align}\label{e:cubic_CS_matching}
 \sum_{R} F(R)\sum_{w \in R}\,w_{I} w_{J} w_{K} = 0 \quad \forall \, I,J,K \, ,
\end{align}
which appear in \eqref{e:non_Abelian_anomaly_4d}, are, depending on the choice of indices, either trivially fulfilled or equivalent to the cancelation
of pure non-Abelian anomalies \eqref{4d_nA_anomaly}
\begin{align}\label{e:cubic_anomaly}
 \sum_R F(R) \, V_R = 0 \, ,
\end{align}
where $V_R$ appears in the trace reduction
\begin{align}\label{e:cubic_trace}
 \tr_R \hat F^3 = V_R\, \tr_f \hat F^3 \, .
\end{align}
Expanding the traces we can write \eqref{e:cubic_trace} as
\begin{align}
 F^I F^J F^K \sum_{w \in R} w_I w_J w_K = F^I F^J F^K \, V_R \, \sum_{w^f} w_I^f w_J^f w_K^f \, ,
\end{align}
where $F=F^I T_I$ and we sum over all weights, in particular $w^f$ denote the weights of the fundamental representation.
Considering this equation as a generating function we find
\begin{align}\label{e:cubic_weight_sum}
 \sum_{w \in R} w_I w_J w_K = V_R \, \sum_{w^f} w_I^f w_J^f w_K^f \, .
\end{align}
The key point is now to try to generally evaluate the sum over the fundamental weights on the right hand side.
This procedure indeed will allow us to relate the factor $V_R$ to a certain sum over the weights in the representation $R$, which appears
in the calculation of one-loop Chern-Simons terms. In the following we carry this out for all simple Lie algebras.
\begin{enumerate}
 \item $\bf A_1$, $\bf B_n$, $\bf C_n$, $\bf D_n$, $\bf E_6$, $\bf E_7$, $\bf E_8$, $\bf F_4$, $\bf G_2$
 
 For these algebras there exists no cubic Casimir operator which is why non-Abelian anomalies are always trivially absent and one
 therefore defines $V_R = 0$. Via \eqref{e:cubic_weight_sum} the condition from the one-loop Chern-Simons matching
 \eqref{e:cubic_CS_matching} is then equivalent to the non-Abelian anomaly cancelation condition \eqref{e:cubic_anomaly}.\footnote{Note also
 that the condition $\sum_{w^f} w_I^f w_J^f w_K^f = 0 \, \, \, \, \forall \, I,J,K$ precisely means that
 there is no cubic Casimir and one then also has by definition $V_R =0$.}
 
 \item $\bf A_{n \neq 1}$
 
 For $A_{n \neq 1}$ there exists a cubic Casimir and we start by
 explicitly evaluating the traces over the fundamental weights for different index choices.
 
 \begin{enumerate}[label=(\alph*)]
 \item \underline{$I=J=K$}
 
 We calculate
 \begin{align}
  \sum_{w^f} ( w_I^f )^3 = 0 \, 
 \end{align}
 such that we can conclude using \eqref{e:cubic_weight_sum}
 \begin{align}
  \sum_{w \in R} ( w_I )^3 = 0 \, .
 \end{align}
 The corresponding Chern-Simons matching \eqref{e:cubic_CS_matching} is therefore trivial and imposes no restrictions on the spectrum.
 
 \item \underline{$I=K\neq J$}
 
 Now we evaluate
 \begin{align}
  \sum_{w^f} ( w_I^f )^2 w_J^f = (I-J)\, \cC_{IJ} \, \, .
 \end{align}
 With \eqref{e:cubic_weight_sum} the Chern-Simons matching \eqref{e:cubic_CS_matching} in this case becomes
 \begin{align}
  \sum_{R} F(R) V_R \, (I-J)\, \cC_{IJ} = 0 \, ,
 \end{align}
 which is equivalent to the anomaly condition \eqref{e:cubic_anomaly}.
 
 \item \underline{$I\neq J \neq K$}
 
 Finally it turns out that
 \begin{align}
  \sum_{w^f} w_I^f w_J^f w_K^f = 0 \, ,
 \end{align}
 which is why the Chern-Simons matching is again trivial like in the case $I=J=K$.
 \end{enumerate}
\end{enumerate}
To put it in a nutshell we have shown that the Chern-Simons matching \eqref{e:cubic_CS_matching} is completely equivalent to the anomaly cancelation
conditions \eqref{e:cubic_anomaly} for all simple Lie algebras.

\subsection{Quartic trace identities}
Let us perform the same steps as in the last subsection now for quartic traces.
More precisely we show that the matching of Chern-Simons coefficients \eqref{e:6D_CS_match}
\begin{align}\label{e:6d_CS_match}
   \sum_{R}  F_{\fe}(R) \sum_{w \in R} \ w_{I} w_{J}  w_{K} w_{L} =  - b^{\alpha} b^{\beta} \Omega_{\alpha\beta} \, ( \cC_{I J}
\cC_{KL}
 + \cC_{I K} \cC_{JL} +  \cC_{IL} \cC_{J K})
\end{align}
is equivalent to the six-dimensional pure non-Abelian anomalies \eqref{e:6d_anom_3}, \eqref{e:6d_anom_4}
\begin{align}\label{e:6d_anomalies_nA}
 \sum_R F_{\fe}(R) B_R &= 0 \, , \\
  \sum_R F_{\fe}(R) C_R  &= -3 \frac{b^\alpha}{\lambda (\mathfrak{g})} \frac{b^\beta}{\lambda (\mathfrak{g})} \Omega_{\alpha\beta}\, , \nn
\end{align}
where the constants $B_R , C_R$ are defined as
\begin{align}\label{e:quartic_trace_red}
 \tr_R \hat F^4 = B_R \, \tr_f \hat F^4 + C_R \, (\tr_f \hat F^2)^2 \, .
\end{align}
Expanding the traces on both sides of \eqref{e:quartic_trace_red} and taking derivatives with respect to $F^I$
we obtain in analogy to \eqref{e:cubic_weight_sum}
\begin{align}\label{e:quartic_trace_weight_sum}
 &\sum_{w \in R} w_I w_J w_K w_L \nn \\
 &\,\,\,\,\, = B_R \sum_{w^f} w^f_I w^f_J w^f_K w^f_L  
  +\frac{1}{3}C_R \bigg [ \Big(\sum_{w^f} w_I^f w_J^f\Big ) \Big(\sum_{ w^{' f}} w_K^{' f} w_L^{' f}\Big ) + \Big(\sum_{w^f} w_I^f w_K^f\Big )
  \Big(\sum_{w^{' f}} w_J^{' f} w_L^{' f}\Big ) \nn\\
 &\hspace{175pt}+ \Big(\sum_{w^f} w_I^f w_L^f\Big ) \Big(\sum_{w^{' f}} w_J^{' f} w_K^{' f}\Big ) \bigg ] \, .
\end{align}

Like in the preceding subsection we now evaluate explicitly the sums over the fundamental weights in order to rewrite \eqref{e:quartic_trace_weight_sum}.
For the different simple Lie algebras and all possible choices of indices \eqref{e:quartic_trace_weight_sum} then becomes:
\begin{enumerate}
 \item $\bf A_n$

 \begin{enumerate}
	\item \underline{$I=J=K=L$}
	\begin{align}
	 \sum_{w \in R} (w_I)^4 = B_R \,\, \mathcal{C}_{II}\,\, \lambda (\mathfrak{g}) + C_R \,\, \mathcal{C}_{II}^2 \,\, \lambda (\mathfrak{g})^2 \, ,
	\end{align}
	\item \underline{$I=K=L$, $I \neq J$}
	\begin{align}
	 \sum_{w \in R} (w_I)^3 \,  w_J = B_R \,\, \mathcal{C}_{IJ}\,\, \lambda (\mathfrak{g}) + C_R  \,\,
	 \mathcal{C}_{II} \,\,\mathcal{C}_{IJ} \,\, \lambda (\mathfrak{g})^2  \, ,
	\end{align}
	\item \underline{$I=L$, $I \neq J \neq K$}
	\begin{align}
	 \sum_{w \in R} (w_I)^2 \,  w_J w_K = \frac{1}{3}\,\,C_R \,\big ( 2\,\, \mathcal{C}_{IJ} \,\,\mathcal{C}_{IK} + \mathcal{C}_{II} \,\,\mathcal{C}_{JK} \big )
	 \,\, \lambda (\mathfrak{g})^2  \, ,
	\end{align}
	\item \underline{$I \neq J \neq K \neq L$}
	\begin{align}
	 \sum_{w \in R} w_I w_J w_K w_L = \frac{1}{3}\,\,C_R \,\big ( \mathcal{C}_{IJ} \,\,\mathcal{C}_{KL} +  \mathcal{C}_{IK} \,\,\mathcal{C}_{JL} +  \mathcal{C}_{IL} \,\,\mathcal{C}_{JK}  \big )
	 \,\, \lambda (\mathfrak{g})^2  \, .
	\end{align}
 \end{enumerate}
 We can now insert these equations into the Chern-Simons matching \eqref{e:6d_CS_match} and find two linearly independent equations
 \begin{align}
  \sum_R F_{\fe}(R) \bigg(\frac{1}{2}\,B_R + C_R \bigg)  &= -3\, \frac{b^\alpha}{\lambda(\mathfrak{g})}
  \,\,\frac{b^\beta}{\lambda(\mathfrak{g})} \Omega_{\alpha\beta} \, , \\
   \sum_R F_{\fe}(R)\,\, C_R   &= -3\, \frac{b^\alpha}{\lambda(\mathfrak{g})}
  \,\,\frac{b^\beta}{\lambda(\mathfrak{g})} \Omega_{\alpha\beta} \, . \nn 
 \end{align}
These equations are in fact equivalent to the gauge anomaly conditions \eqref{e:6d_anomalies_nA}.
 
 \item $\bf B_n$
 
 \begin{enumerate}
	\item \underline{$I=J=K=L$}
	\begin{align}
	 \sum_{w \in R} (w_I)^4 = \frac{1}{4}\,B_R \,\,\mathcal{C}_{In}^2\,\, \mathcal{C}_{II}\,\, \lambda (\mathfrak{g})
	 + C_R \,\, \mathcal{C}_{II}^2 \,\, \lambda (\mathfrak{g})^2  \, ,
	\end{align}
	\item \underline{$I=K=L$, $I \neq J$}
	\begin{align}
	 \sum_{w \in R} (w_I)^3 \,  w_J = \frac{1}{4}\,B_R \,\,\mathcal{C}_{In}^2\,\,\mathcal{C}_{IJ}\,\, \lambda (\mathfrak{g}) + C_R  \,\,
	 \mathcal{C}_{II} \,\,\mathcal{C}_{IJ} \,\, \lambda (\mathfrak{g})^2  \, ,
	\end{align}
	\item \underline{$I=L$, $I \neq J \neq K$}
	\begin{align}
	 \sum_{w \in R} (w_I)^2 \,  w_J w_K = 
	 \frac{1}{3}\,\,C_R \,\big ( 2\,\, \mathcal{C}_{IJ} \,\,\mathcal{C}_{IK} + \mathcal{C}_{II} \,\,\mathcal{C}_{JK} \big )
	 \,\, \lambda (\mathfrak{g})^2  \, ,
	\end{align}
	\item \underline{$I \neq J \neq K \neq L$}
	\begin{align}
	 \sum_{w \in R} w_I w_J w_K w_L = \frac{1}{3}\,\,C_R \,\big ( \mathcal{C}_{IJ} \,\,\mathcal{C}_{KL} +  \mathcal{C}_{IK} \,\,\mathcal{C}_{JL} +  \mathcal{C}_{IL} \,\,\mathcal{C}_{JK}  \big )
	 \,\, \lambda (\mathfrak{g})^2  \, .
	\end{align}
 \end{enumerate}
 Insertion into \eqref{e:6d_CS_match} yields
 \begin{align}
  \sum_R F_{\fe}(R) \bigg(\frac{1}{4}\,B_R + C_R \bigg)  &= -3\, \frac{b^\alpha}{\lambda(\mathfrak{g})}
  \,\,\frac{b^\beta}{\lambda(\mathfrak{g})} \Omega_{\alpha\beta} \, , \\
   \sum_R F_{\fe}(R)\,\, C_R   &= -3\, \frac{b^\alpha}{\lambda(\mathfrak{g})}
  \,\,\frac{b^\beta}{\lambda(\mathfrak{g})} \Omega_{\alpha\beta} \, , \nn 
 \end{align}
which is equivalent to \eqref{e:6d_anomalies_nA}.
 
 \item $\bf C_n$
 
 \begin{enumerate}
	\item \underline{$I=J=K=L$}
	\begin{align}
	 \sum_{w \in R} (w_I)^4 = B_R \,\, \mathcal{C}_{II}\,\, \lambda (\mathfrak{g}) + C_R \,\, \mathcal{C}_{II}^2 \,\, \lambda (\mathfrak{g})^2  \, ,
	\end{align}
	\item \underline{$I=K=L$, $I \neq J$}
	\begin{align}
	 \sum_{w \in R} (w_I)^3 \,  w_J = B_R \,\, \mathcal{C}_{IJ}\,\, \lambda (\mathfrak{g}) + C_R  \,\,
	 \mathcal{C}_{II} \,\,\mathcal{C}_{IJ} \,\, \lambda (\mathfrak{g})^2 \, ,
	\end{align}
	\item \underline{$I=L$, $I \neq J \neq K$}
	\begin{align}
	 \sum_{w \in R} (w_I)^2 \,  w_J w_K = 
	 \frac{1}{3}\,\,C_R \,\big ( 2\,\, \mathcal{C}_{IJ} \,\,\mathcal{C}_{IK} + \mathcal{C}_{II} \,\,\mathcal{C}_{JK} \big )
	 \,\, \lambda (\mathfrak{g})^2  \, ,
	\end{align}
	\item \underline{$I \neq J \neq K \neq L$}
	\begin{align}
	 \sum_{w \in R} w_I w_J w_K w_L = \frac{1}{3}\,\,C_R \,\big ( \mathcal{C}_{IJ} \,\,\mathcal{C}_{KL} +  \mathcal{C}_{IK} \,\,\mathcal{C}_{JL} +  \mathcal{C}_{IL} \,\,\mathcal{C}_{JK}  \big )
	 \,\, \lambda (\mathfrak{g})^2  \, ,
	\end{align}
 \end{enumerate}
 which can be inserted into \eqref{e:6d_CS_match}
 \begin{align}
  \sum_R F_{\fe}(R) \bigg(\frac{1}{4}\,B_R + C_R \bigg)  &= -3\, \frac{b^\alpha}{\lambda(\mathfrak{g})}
  \,\,\frac{b^\beta}{\lambda(\mathfrak{g})} \Omega_{\alpha\beta} \, , \\
   \sum_R F_{\fe}(R)\,\, C_R   &= -3\, \frac{b^\alpha}{\lambda(\mathfrak{g})}
  \,\,\frac{b^\beta}{\lambda(\mathfrak{g})} \Omega_{\alpha\beta} \, . \nn 
 \end{align}
These equations are equivalent to the anomaly conditions \eqref{e:6d_anomalies_nA}.

 \item $\bf D_n$
 
 \begin{enumerate}
	\item \underline{$I=J=K=L$}
	\begin{align}
	 \sum_{w \in R} (w_I)^4 = B_R \,\, \mathcal{C}_{II}\,\, \lambda (\mathfrak{g}) + C_R \,\, \mathcal{C}_{II}^2 \,\, \lambda (\mathfrak{g})^2  \, ,
	\end{align}
	\item \underline{$I=K=L$, $I \neq J$}
	\begin{align}
	 \sum_{w \in R} (w_I)^3 \,  w_J = B_R \,\, \mathcal{C}_{IJ}\,\, \lambda (\mathfrak{g}) + C_R  \,\,
	 \mathcal{C}_{II} \,\,\mathcal{C}_{IJ} \,\, \lambda (\mathfrak{g})^2 \, ,
	\end{align}
	\item \underline{$I=L$, $I \neq J \neq K$}
	\begin{align}
	 \sum_{w \in R} (w_I)^2 \,  w_J w_K = \alpha_{IJK}\,B_R +
	 \frac{1}{3}\,\,C_R \,\big ( 2\,\, \mathcal{C}_{IJ} \,\,\mathcal{C}_{IK} + \mathcal{C}_{II} \,\,\mathcal{C}_{JK} \big )
	 \,\, \lambda (\mathfrak{g})^2  \, ,
	\end{align}
	\item \underline{$I \neq J \neq K \neq L$}
	\begin{align}
	 \sum_{w \in R} w_I w_J w_K w_L = \frac{1}{3}\,\,C_R \,\big ( \mathcal{C}_{IJ} \,\,\mathcal{C}_{KL} +  \mathcal{C}_{IK} \,\,\mathcal{C}_{JL} +  \mathcal{C}_{IL} \,\,\mathcal{C}_{JK}  \big )
	 \,\, \lambda (\mathfrak{g})^2  \, ,
	\end{align}
 \end{enumerate}
 with the definition
 \begin{align}
  \alpha_{IJK} := 4\,\Big(\delta_{I,n-2}\,\delta_{(J,n}\,\delta_{K),n-1}-\delta_{I,n-1}\,\delta_{(J,n}\,\delta_{K),n-2}-\delta_{I,n}\,
 \delta_{(J,n-1}\,\delta_{K),n-2}\Big)
 \end{align}
 Inserting into \eqref{e:6d_CS_match} we obtain
 \begin{align}
  \sum_R F_{\fe}(R) \bigg(\frac{1}{4}\,B_R + C_R \bigg)  &= -3\, \frac{b^\alpha}{\lambda(\mathfrak{g})}
  \,\,\frac{b^\beta}{\lambda(\mathfrak{g})} \Omega_{\alpha\beta} \, , \\
   \sum_R F_{\fe}(R)\,\, C_R   &= -3\, \frac{b^\alpha}{\lambda(\mathfrak{g})}
  \,\,\frac{b^\beta}{\lambda(\mathfrak{g})} \Omega_{\alpha\beta} \, , \nn 
 \end{align}
  which is equivalent to the anomaly conditions \eqref{e:6d_anomalies_nA}.

  \item $\bf E_6$, $\bf E_7$, $\bf E_8$, $\bf F_4$, $\bf G_2$ 
  
  For these algebras there is no fourth-order Casimir, therefore by definition $B_R = 0$ for all representations. We find by explicit calculation
  \begin{enumerate}
	\item \underline{$I=J=K=L$}
	\begin{align}
	 \sum_{w \in R} (w_I)^4 =  C_R \,\, \mathcal{C}_{II}^2 \,\, \lambda (\mathfrak{g})^2 
	\end{align}
	\item \underline{$I=K=L$, $I \neq J$}
	\begin{align}
	 \sum_{w \in R} (w_I)^3 \,  w_J =  C_R  \,\, \mathcal{C}_{II} \,\,\mathcal{C}_{IJ} \,\, \lambda (\mathfrak{g})^2 
	\end{align}
	\item \underline{$I=L$, $I \neq J \neq K$}
	\begin{align}
	 \sum_{w \in R} (w_I)^2 \,  w_J w_K = \frac{1}{3}\,\,C_R \,\big ( 2\,\, \mathcal{C}_{IJ} \,\,\mathcal{C}_{IK} + \mathcal{C}_{II} \,\,\mathcal{C}_{JK} \big )
	 \,\, \lambda (\mathfrak{g})^2 
	\end{align}
	\item \underline{$I \neq J \neq K \neq L$}
	\begin{align}
	 \sum_{w \in R} w_I w_J w_K w_L = \frac{1}{3}\,\,C_R \,\big ( \mathcal{C}_{IJ} \,\,\mathcal{C}_{KL} +  \mathcal{C}_{IK} \,\,\mathcal{C}_{JL} +  \mathcal{C}_{IL} \,\,\mathcal{C}_{JK}  \big )
	 \,\, \lambda (\mathfrak{g})^2 
	\end{align}
 \end{enumerate}
 Plugging this in into the Chern-Simons matching \eqref{e:6d_CS_match} we get
 \begin{align}
   \sum_R F_{\fe}(R)\,\, C_R   = -3\, \frac{b^\alpha}{\lambda(\mathfrak{g})}
  \,\,\frac{b^\beta}{\lambda(\mathfrak{g})} \Omega_{\alpha\beta} \, ,
 \end{align}
which is again equivalent to the cancelation of anomalies since the first equation in \eqref{e:6d_anomalies_nA} is trivial due to the absence of
a fourth-order Casimir.
\end{enumerate}

Thus we have shown that the matching condition from one-loop Chern-Simons terms \eqref{e:6d_CS_match} is fully equivalent
to the cancelation of non-Abelian gauge anomalies \eqref{e:6d_anomalies_nA} for all simple Lie algebras.

\section{Intersection numbers and matchings}\label{sec:intersection_nos}

In this section we list useful intersection numbers of elliptically fibered Calabi-Yau four- and threefolds along with their matched
quantity in the M-/F-theory duality. The definitions of the indices are as in \autoref{sec:F-theory}.

For Calabi-Yau fourfolds we consider 
\begin{align}
 \Theta_{\Lambda\Sigma} = - \frac{1}{4}\ D_\Lambda \cdot D_\Sigma \cdot G_4 \, , \qquad
 \tensor{\cK}{_\Lambda_\Sigma^\alpha} = \tensor{\eta}{^{-1}_\beta^\alpha}\ D_\Lambda \cdot D_\Sigma \cdot \cC^\beta \, ,
\end{align}
where $\tensor{\eta}{_\alpha^\beta}$ is the full-rank intersection matrix \eqref{e:def_metric}.
One finds
\begin{align}
  \Theta_{\alpha\beta} &= 0 \, ,  &\Theta_{\alpha 0} &= 0 \, , 
 &\Theta_{\alpha I} &= 0 \, ,  &\Theta_{\alpha m} & = \frac{1}{2}\theta_{\alpha m} \, , \nn \\
 \tensor{\cK}{_\alpha_\beta^\gamma} & = 0 \, ,  &\tensor{\cK}{_0_\alpha^\beta} & = \delta^\beta_\alpha \, , 
 &\tensor{\cK}{_m_\alpha^\beta} & = 0 \, ,  & \tensor{\cK}{_I_\alpha^\beta} & = 0 \, , \nn \\
 \tensor{\cK}{_0_0^\alpha} & = 0 \, ,  &\tensor{\cK}{_0_m^\alpha} & = 0 \, , 
  &\tensor{\cK}{_0_I^\alpha} & = 0 \, , &\tensor{\cK}{_m_n^\alpha} & = - b_{mn}^\alpha \, , \nn \\
   &&&&&&\tensor{\cK}{_I_J^\alpha} & = - b^\alpha \ \cC_{IJ} \, .
\end{align}

Finally for Calabi-Yau threefolds there are intersection numbers
\begin{align}
 k_{\Lambda\Sigma\Theta} = D_\Lambda \cdot D_\Sigma \cdot D_\Theta \, , \qquad
 k_{\Lambda} = D_\Lambda \cdot c_2 \, ,
\end{align}
and one evaluates
\begin{align}
 &k_{\alpha\beta\gamma} =0 \ , &&k_{ 0\alpha\beta} = \Omega_{\alpha\beta} \ , &&k_{I\alpha\beta} =0 \ , \nn \\
 &k_{m\alpha\beta} =0 \ , && k_{00\alpha} =0 \ , && k_{IJ\alpha} = -\Omega_{\alpha\beta} b^\beta \ \cC_{IJ} \ , \nn \\
 &k_{mn\alpha} = -\Omega_{\alpha\beta} b^\beta_{mn} \ , &&k_{0I\alpha} =0 \ , &&k_{0m\alpha} =0 \ ,\nn \\ 
 &k_{\alpha} = -12\ \Omega_{\alpha\beta} a^\beta \ .
\end{align}

\bibliography{references}
\bibliographystyle{utcaps} 
\end{document}